\documentclass[prb,twocolumn,preprintnumbers,superscriptaddress]{revtex4-2}

\usepackage{amsmath}
\usepackage{xcolor}
\usepackage{graphicx}
\usepackage{dcolumn}
\usepackage{bm}
\usepackage{hyperref}

\newcommand{\vect}[1]{\boldsymbol{\mathbf{#1}}}
\DeclareMathOperator{\e}{e}

\begin{document}


\title{Stimulated down-conversion of single-photon emission in a quantum dot placed in a target-frequency microcavity}

\author{I.V.~Krainov}
\email{igor.kraynov@mail.ru} 
\affiliation{Ioffe Institute, 194021 St.~Petersburg, Russia} 
\author{M.V.~Rakhlin}
\affiliation{Ioffe Institute, 194021 St.~Petersburg, Russia} 
\author{A.I.~Veretennikov}
\affiliation{Ioffe Institute, 194021 St.~Petersburg, Russia} 
\author{T.V.~Shubina}
\affiliation{Ioffe Institute, 194021 St.~Petersburg, Russia}

\date{\today}

\begin{abstract}
Currently, two optical processes are mainly used to realize single photon sources: deterministic transitions in a semiconductor quantum dot (QD) placed in a microcavity and spontaneous frequency down-conversion in materials with intrinsic nonlinearity. In this work, we consider another approach that combines the advantages of both, such as high power with on-demand generation from QDs and the possibility of frequency tuning from nonlinear sources. For this purpose, we use stimulated frequency down-conversion occurring directly in the QD inside a microcavity designed not to the exciton frequency in the QD but to the target single photon frequency, which is set by the difference between the exciton resonance and the stimulating laser energies. This  down-conversion arises from the second-order nonlinear interaction of an exciton (bright heavy-hole or dark) and a light-hole exciton in the stimulating laser field. We present an analytical model for such a down-conversion process and evaluate its efficiency for a widely sought-after single photon source for the telecom C-band (1530-1565 nm). 
We show that the emission rate of down-converted single photons can approach MHz. At certain conditions, this process is comparable in efficiency to direct emission from an InAs/GaAs QD at 920 nm, which is outside the cavity mode.
\end{abstract}


\maketitle

\section{Introduction}
Modern optical quantum technologies such as quantum communication and quantum data processing require efficient sources that emit single photons with high purity and indistinguishability. Two optical processes are currently being developed to realize such single-photon sources:  i)  on-demand deterministic transition between the excited and ground exciton states  in a semiconductor quantum dot (QD) placed in a microcavity \cite{Senellart2017, Rakhlin2023, Vajner2022} and ii) spontaneous parametric frequency down-conversion (SPDC) or spontaneous four-wave mixing (SFWM) in different media with second- or third-order susceptibilities, respectively \cite{Caspani2017, Baboux2023, Wang2024}. 

For the first trend, the most widely used are InAs/GaAs QDs placed in microresonators, which have undeniable advantages such as compactness, high single-photon generation efficiency (end-to-end efficiency $\sim$0.6), spontaneous emission rate above MHz (GHz is the limit), indistinguishability and single-photon purity close to unity \cite{Wang2019,Tomm2021}. The recently reported efficiency of 0.7 is essentially the threshold sufficient for scalable photonic quantum computing \cite{Ding2023}.
However, the InAs/GaAs QDs emit efficiently only in the wavelength range of 900–1000 nm.  Mastering other ranges, such as those corresponding to telecom frequencies, still remains a challenge for widespread application, despite some promising results  \cite{Nawrath2023}. 
Another serious problem is that QDs produced by epitaxial methods are characterized by a significant spread of emission energy, while the fundamental mode in microresonators has a well-reproducible energy. As a result, it is difficult to ensure precise coupling QD emission and the fundamental mode to take advantages of the Purcell enhancement and efficient light yield. The ability to form several identical single-photon sources requires laborious testing of numerous samples or the use of adjusting piezoelectric units \cite{Yang2024}.

In contrast, single-photon sources produced using nonlinear SPDC and SFWM processes can have very constant energy. The flexibility in tuning the single photon generation energy, strictly related to the pump laser energy, makes these sources compatible with opto-telecommunication systems in the C-band (1530-1565 nm) \cite{Paesani2020}. These spontaneous nonlinear processes are characterized by low efficiency, which depends on the pump field intensity, which cannot be increased indefinitely, since this would destroy the purity of single photons. Although the second-order process SPDC provides higher efficiency than SFWM, for it is more difficult to satisfy the momentum conservation rule \cite{Helmy2011}. 
In both spontaneous processes, pairs of single photons satisfying the energy conservation relation appear non-deterministically, at random times. However, both sources are fully compatible with chip technology, which allows for the widespread use of low-power source multiplexing with heralding \cite{Ma2011, Scott2020}. To implement this principle, the photons in a pair are separated, and one of them can then be used to herald that the other --- a signal photon --- is ready to be delivered to the working circuit \cite{Spring2013}. In addition, this power problem is addressed by increasing the planar dimensions and/or using optical microresonators of various designs \cite{Wang2021,Wang2024}. However, all such measures increase the size of the single photon sources and require numerous electro-optical modulators, the advanced chip technology for which is still under development \cite{Eltes2020}.

A hybrid system has been proposed to combine the advantages of these two approaches, in which an InAs/GaAs QD in a microcavity generates single photons, while a coupled nonlinear element such as a periodically polarized lithium niobate (ppLN) waveguide provides their nonlinear down-conversion under stimulated laser pumping \cite{Pelc2012, Kambs2016, Lio2022, Morrison2021}. Such experiments have mainly focused on achieving efficient single-photon generation in the telecom C-band. The down-converted photons, generated on demand in a second-order nonlinear process, retain their quantum-statistical properties \cite{Kambs2016}. The best values obtained were g$_2(0)$=0.024 and indistinguishability of 0.948 \cite{Lio2022}. However, some degradation of these parameters relative to the original QD radiation was recognized \cite{Singh2019}. The conversion efficiency, depending on the power of the stimulating laser, can reach 50$\%$ of the QD radiation, but when pumped by a 150 mW stimulating laser \cite{Morrison2021}. Note that this power is several orders of magnitude higher than with direct generation of single QD photons.

In this paper, we propose another approach that also combines the advantages of the two main trends, namely, quantum stimulated down-conversion (SDC) of single photons, which occurs directly in a QD placed in a microcavity specifically designed for the target single photon frequency but not for the QD exciton resonance. The process exploits the second-order nonlinear interaction between a light hole exciton and a recombining bright exciton (heavy hole exciton) or dark exciton in a stimulating laser field. The target single photon emission frequency is set by the difference between the exciton and stimulating laser energies. This method provides high tunability of the single photon frequency. As a proof of concept, we discuss the implementation of a sought-after 1550 nm single photon source using a QD emitting at 920 nm.

\section{Concept of quantum SDC in a single QD}

\begin{figure}
\includegraphics[width=1\linewidth]{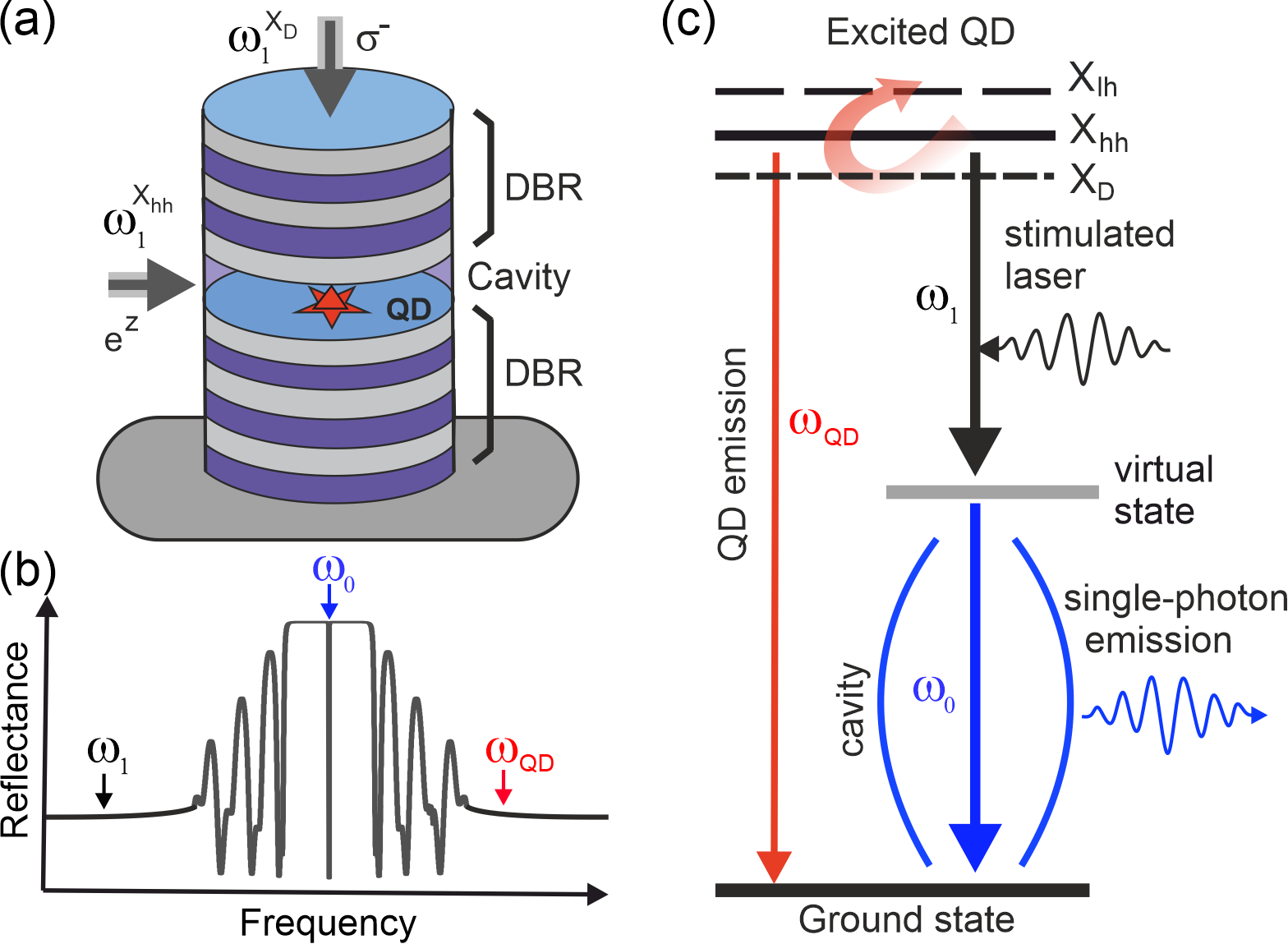}
\centering
\caption{\label{figStructure} 
(a) Single photon source based on a QD inside a target-frequency microcavity. Arrows indicate the direction of stimulating excitation for bright $X_{hh}$ and dark $X_D$ excitons. The needed linear $\e^z$ or circular $\sigma^-$ polarizations are specified.  (b) Reflectance spectrum of the Bragg microcavity calculated for emission in telecom range. Arrows indicate frequencies of stimulating laser $\omega_{1}$, cavity mode $\omega_{0}$ and QD exciton $\omega_{QD}$ radiation. (c) Schematic of single photon generation during down-conversion of the  frequency $\omega_{QD}$ to difference frequency $\omega_{0}$ under laser pumping with frequency $\omega_{1}$ ($\omega_{0}$=$\omega_{QD}$-$\omega_{1}$). The process is caused by second-order nonlinearity arising from interaction with excited state of light-hole exciton.}
\label{figDBR}
\end{figure}   

The idea of frequency down-conversion in an object having quantum levels has a long history, initiated by studies of stimulated Raman scattering in potassium atoms \cite{Yatsiv1968} and then in Rb atoms with a $\Lambda$-level scheme placed in an optical cavity \cite{Hennrich2000}. More recently, some frequency tuning and its dependence on the stimulating laser pulse have been studied in a quantum three-level system such as a biexciton ladder in QDs \cite{Schumacher2016, Schumacher2022}. Recently, a laser-controlled nonlinear process on the biexciton ladder has been studied in InAs QDs without a cavity \cite{Jonas2022} and with a cavity designed for the excitonic resonance in the QD to obtain a distinct polarization of the single-photon emission \cite{Wei2022}. In particular, Wei et al. obtained a single-photon purity of 0.998 and an indistinguishability of 0.926, but the source brightness was only 3\% \cite{Wei2022}.
Thus, the previous experiments confirmed the possibility of obtaining single photons with good statistical characteristics in a nonlinear process occurring directly in the InAs/GaAs QD. The goal of approaching the telecommunication C-band and enhancing the brightness of single-photon radiation, as in our work, was not set.

The proposed scheme for implementing efficient quantum SDC in the QD-resonator system is shown in Fig. 1a. An important feature is that the QD is fabricated using a well-established technology for the 900-950 nm range. The QD is located inside a microresonator with distributed Bragg reflectors (DBRs) designed for a target frequency different from the QD exciton resonance frequency, but equal to the difference between that and the stimulating laser frequency, as shown in the calculated microresonator reflectance spectrum in Fig. 1b. In such a system, the frequency down-conversion is a second-order nonlinear process requiring the presence of another excited level in the QD, which in our case is the light hole exciton level (X$_{lh}$). The nonlinear interaction of a heavy hole exciton (X$_{hh}$) or a dark exciton (X$_{D}$), if it is properly prepared, with X$_{lh}$ in the strong field of the stimulating laser results in the appearance of a virtual level through which recombination occurs, leading to single-photon emission of the difference frequency (Fig. 1c).

To ensure such a nonlinear interaction, the excitation methods must be different for the bright and dark exciton states due to symmetry restrictions. Namely, the coupling of X$_{hh}$ with X$_{lh}$ is possible only with excitation incident on the facet of the columnar microresonator, whereas for a dark exciton this can be done with excitation at its top. The same applies to the required light polarization, which must be either linear or circular, as shown in Fig. 1(a).

\section{Theoretical consideration}

Let us consider the QD with dark and bright exciton levels which can be initialized in the states:
\begin{gather}
\label{psiDE}
    \left|\psi_{\text{XD}}\right\rangle = \frac{\left| +2 \right\rangle - \left| -2 \right\rangle}{\sqrt{2}}, \\
    \left|\psi_{\text{Xhh}}\right\rangle = \left| -1 \right\rangle.
    \label{psiBE}
\end{gather}
The initialization of dark excitons and their parameters was investigated in \cite{Schawrtz2015}. 
From the quantum mechanical point of view interaction with electromagnetic field leads to the transitions between electronic states via Hamiltonian $ (e/mc) \vect{A} \hat{\vect{p}}$. This interaction leads to transitions from initial state $\left|i\right\rangle$ to the all other states $\left|f\right\rangle$ with nonzero momentum matrix element $p_{fi}$, in zero order on $1/\lambda$ - inverse wavelength. On the times much more than $\hbar / (E_f - E_i)$ transitions which satisfy energy conservation $E_f - E_i = \hbar \omega$ will be active, where $\omega$ - nonzero harmonic of electromagnetic field. 
If the electromagnetic field contains multiple harmonics (the interaction with which is not so small), then in the second order of perturbation theory indirect transitions also occur through intermediate states $\left|k\right\rangle$. Energy conservation will only be valid between the initial and final states, for example $E_f - E_i = \hbar \omega_0 + \hbar \omega_1$, where $\omega_0$ will be the cavity mode, and $\omega_1$ will be the exciting beam in our case. By increasing the power of such an exciting beam, we can make this process observable, which will be discussed at the end of this article.

\begin{figure} 
\includegraphics[width=1\linewidth]{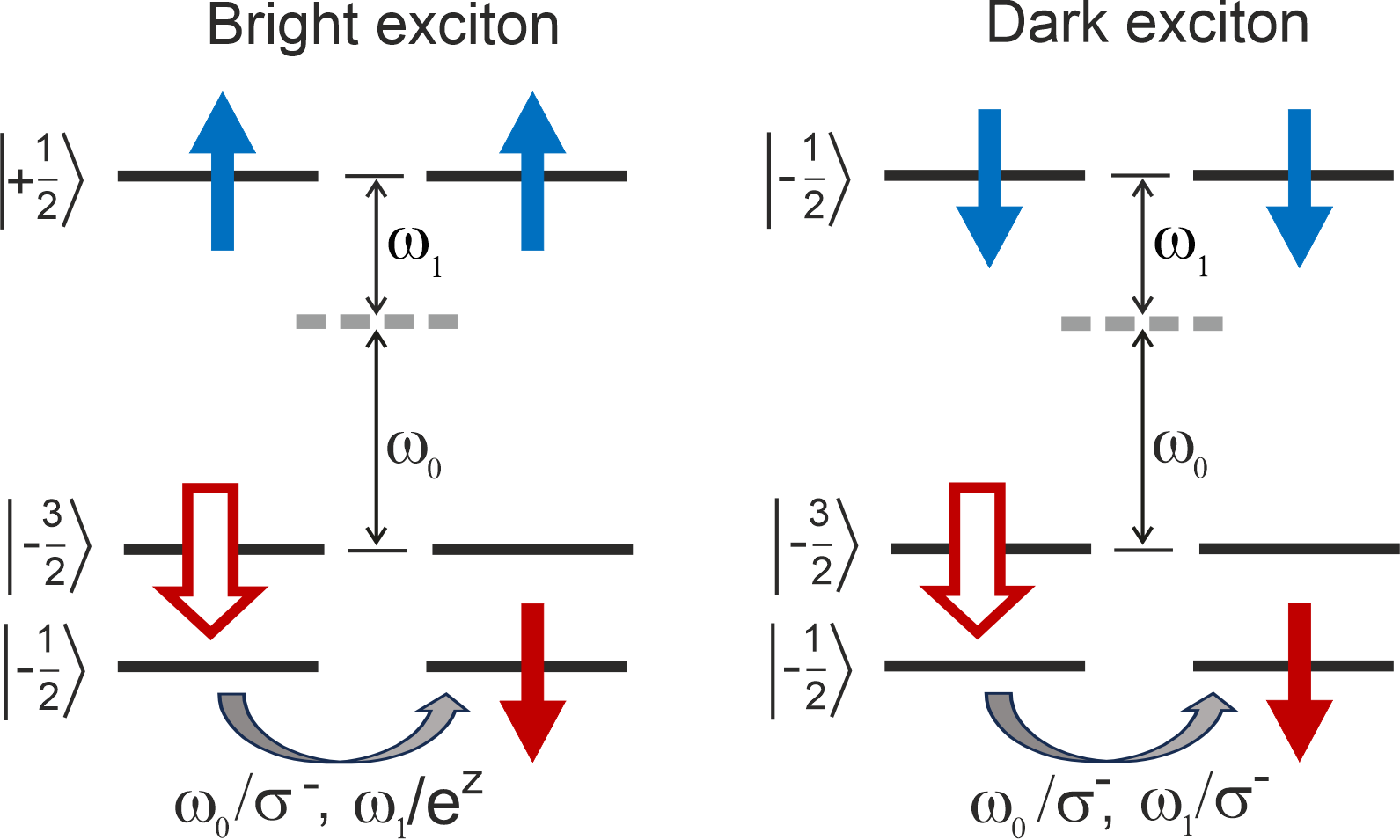}
\centering
\caption{\label{figStructure} 
Stimulated frequency down-conversion schemes for bright and dark excitons in a QD, whose emission is subjected to a $\omega_1$ laser field and coupled finally to a $\omega_0$ cavity mode. Desired linear $\e^z$ or circular  $\sigma^-$ polarizations are indicated.  
}
\end{figure}
To describe the SDC from the states given by equations (\ref{psiDE}) and (\ref{psiBE}), it is necessary to consider optical transitions between the initial, intermediate and final states. The most intense recombination channels in such a nonlinear process should have optically allowed transitions at all stages. Below we show that light holes are an ideal candidate for this. We assume that a standard rotationally symmetric column resonator  is grown along the $z$ direction with an optical cavity that has a fundamental mode $\omega_0$ and is doubly degenerate in the $x, y$ polarizations. The optical transitions between electronic states are determined by the matrix element $ (e/mc) \vect{A} \hat{\vect{p}}$. The most intense of them have nonzero momenta between the Bloch part of the wave function, i.e., optically allowed transitions. The Bloch wave functions of the conduction and valence band states, including light holes, are as follows:
\begin{align*}
    \Gamma_6 (+ 1/2): & \quad  s|\uparrow\rangle, \\
    \Gamma_6 (- 1/2): & \quad  s|\downarrow\rangle, \\
    \Gamma_8 (-3/2): & \quad  -\frac{X+ i Y}{\sqrt{2}}|\uparrow\rangle, \\
    \Gamma_8 (-1/2): & \quad \sqrt{\frac{2}{3}} Z|\uparrow\rangle - \frac{X+ i Y}{\sqrt{6}}|\downarrow\rangle, \\
    \Gamma_8 (+1/2): & \quad  \sqrt{\frac{2}{3}} Z|\downarrow\rangle + \frac{X- i Y}{\sqrt{6}}|\uparrow\rangle, \\
    \Gamma_8 (+3/2): & \quad  \frac{X- i Y}{\sqrt{2}}|\downarrow\rangle.
\end{align*}
The dark exciton has total angular moment $\pm2$ and we need two photons with $\pm1$ spin:
\begin{align}
\nonumber
    \left| +2 \right\rangle = \left|\uparrow_{\Gamma_6}, +3/2_{\Gamma_8}\right\rangle \underset{\sigma_+^1 \, \text{or}\, \sigma_+^0}{\rightarrow} \left|\uparrow_{\Gamma_6}, +1/2_{\Gamma_8} \right\rangle
    \underset{\sigma_+^0 \, \text{or} \, \sigma_+^1}{\rightarrow} \left|0\right\rangle, \\ \nonumber
    \left| -2 \right\rangle = \left|\downarrow_{\Gamma_6}, -3/2_{\Gamma_8}\right\rangle \underset{\sigma_-^1 \, \text{or}\, \sigma_-^0}{\rightarrow} \left|\downarrow_{\Gamma_6}, -1/2_{\Gamma_8} \right\rangle
    \underset{\sigma_-^0 \, \text{or} \, \sigma_-^1}{\rightarrow} \left|0\right\rangle, 
\end{align}
where $\sigma_\pm^{0,1}$ is the photon admitted to the transition with its $\pm$ circular polarization, the superscript specifies the cavity or the stimulated beam. For a bright exciton, we only need to transfer $-1$ spin to the photons, so the stimulated beam must be in $z$ polarization:
\begin{align}
\nonumber
    \left| -1 \right\rangle = \left|\uparrow_{\Gamma_6}, -3/2_{\Gamma_8}\right\rangle \underset{\sigma_-^0}{\rightarrow} \left|\uparrow_{\Gamma_6}, -1/2_{\Gamma_8} \right\rangle
    \underset{\sigma_z^1}{\rightarrow} \left|0\right\rangle.
\end{align}
These processes are described by the following matrix elements:
\begin{align*}
    \left\langle \downarrow_{\Gamma_8} \right| \vect{e}_- \hat{\vect{p}} \left|-3/2_{\Gamma_8} \right\rangle &= \left\langle \uparrow_{\Gamma_8} \right| \vect{e}_+ \hat{\vect{p}} \left|+3/2_{\Gamma_8} \right\rangle = \\ &=\frac{2i}{\sqrt{3}} \langle Z |\hat{p}_y | X \rangle = p_{vv}, \\
    \left\langle \uparrow_{\Gamma_6} \right| \vect{e}_+ \hat{\vect{p}} \left|+1/2_{\Gamma_8} \right\rangle &= -\left\langle \downarrow_{\Gamma_6} \right| \vect{e}_- \hat{\vect{p}} \left|-1/2_{\Gamma_8} \right\rangle =  \\
    &=\sqrt{\frac{2}{3}} \langle s |\hat{p}_x | X \rangle = p_{cv}, \\
    \left\langle \uparrow_{\Gamma_6} \right| \text{e}_z \hat{\text{p}}_z \left|-1/2_{\Gamma_8} \right\rangle &= 
    \left\langle \downarrow_{\Gamma_6} \right| \text{e}_z \hat{\text{p}}_z \left|+1/2_{\Gamma_8} \right\rangle = p_{cv}.
\end{align*}
We assume that all states have $s$-type envelope and their overlapping renormalize $p_{cv}$, $p_{vv}$ matrix elements. Note that $p_{vv}$ is non zero for crystal with inversion assymetry.

We introduce the following operators to describe the SDC dynamics for the  excitons under a stimulated laser beam: $\hat{b}^\dag_{0\pm}$ -- the creation operator of cavity $\sigma^0_\pm$ mode, $\hat{a}^\dag_{2\pm} |0\rangle = \left|\pm1/2_{\Gamma_6}, \pm3/2_{\Gamma_8}\right\rangle$ -- the dark exciton ($\text{X}_\text{D}$) creation operators, $\hat{a}^\dag_{1\pm}  |0\rangle = \left|\mp1/2_{\Gamma_6}, \pm3/2_{\Gamma_8} \right\rangle$ -- the bright exciton ($\text{X}_\text{hh}$) creation operators, $\hat{a}^\dag_{1'\pm}  |0\rangle = \left|\pm1/2_{\Gamma_6}, \pm1/2_{\Gamma_8} \right\rangle$ - the excited  bright exciton ($\text{X}_\text{lh}$)  creation operators. 
The auxiliary stimulated beam is a single-mode monochromatic light containing many photons, and we can consider it classically:
\begin{gather}
    \vect{A}_{1} = \vect{e}_1 \sqrt{\frac{S_p}{k_1 \omega_1}} \cos (\omega_1 t)
\end{gather}
where $\vect{A}_1$ is the stimulated beam electromagnetic vector potential at the QD position and $S_p$ is the absolute value of the Poynting vector here ($S_p \approx P / L^2$ for a laser with power $P$ and focusing in the region $L^2$).

\textbf{Bright exciton.} 
For the initial state of a bright exciton (XB) that is $\text{X}_\text{hh}$ in our case, the polarization of the stimulating beam must be linear: $\vect{e}_1=\vect{e}_z$. The Hamiltonian of the bright exciton dynamics:
\begin{gather} \nonumber
    \hat{H} = \hat{H}_+ + \hat{H}_-, \\ \nonumber
    \hat{H}_+ = (E_1 - i \gamma_{\text{rad}} )\hat{a}^\dag_{1+} \hat{a}_{1+} + 
    (E_{1'} -  i \gamma_{\text{rad}}) \hat{a}^\dag_{1'+} \hat{a}_{1'+} + \\ \nonumber
    +(\omega_0 - i \Gamma) \hat{b}^\dag_{0+} \hat{b}_{0+}  
    +(\hat{b}^\dag_{0+} + \hat{b}_{0+}) (g_0^{vv*} \hat{a}^\dag_{1'+} \hat{a}_{1+} + g_0^{vv} \hat{a}^\dag_{1+} \hat{a}_{1'+})  \\
    +2\cos (\omega_1 t) (g_s^{cv*} \hat{a}^\dag_{1'+} + g_s^{cv} \hat{a}_{1'+}),
\end{gather}
where $E_{1}$ is the energy of the bright exciton, $E_{1'}$ - excited bright  exciton ($\text{X}_\text{lh}$) energy, $\hbar \omega_0$ is the cavity mode energy (for brevity, we assume $\hbar=1$ in the equations here and below), $\Gamma$ is the mode dumping, $g_0^{vv}$ is the coupling constant of exciton states and cavity mode, $g_s^{cv}$ is the coupling constant of the auxiliary laser and the radiating exciton state:
\begin{gather}
    g_s^{cv} = \frac{\sqrt{\alpha \hbar \varepsilon}}{k_0} \frac{p_{cv}}{mc} \sqrt{S_p} ,
\end{gather}
where $\alpha$ is the fine constant, $\varepsilon$ is the permittivity. Note that in the $H_-$ part, in addition to replacing $-$ with $+$ in the operator indices, $p_{cv}$ also changes sign. The direct radiative recombination of the bright exciton state is included in the Hamiltonian via the radiative recombination rate:
\begin{gather}
\label{gammaRad}
    \gamma_{\text{rad}} = \frac{\alpha c \varepsilon k |p_{cv}|^2}{8 (mc)^2 }.
\end{gather}
Other decay and recombination channels can be easily included in this Hamiltonian by redefining $\gamma_{\text{rad}}$ and $\Gamma$. Let the initial state be $\left| -1 \right\rangle=\hat{a}^\dag_{1-} |0\rangle$, whose dynamics is described by the block $\hat{H}_-$. The Hamiltonian $\hat{H}_-$ in the basis of states $(\hat{a}^\dag_{1-} |0\rangle, \hat{b}^\dag_{0-} \hat{a}^\dag_{1'-} |0\rangle, \hat{b}^\dag_{0-} |0\rangle)$ is:
\begin{gather} \nonumber
\hat{H}_- = \\
    \left( \begin{matrix}
        E_1-i\gamma_{\text{rad}} && g_0^{vv} && 0 \\
        g_0^{vv*} && E_{1'} + \omega_0 - i (\gamma_{\text{rad}} + \Gamma) && 2 g_s^{cv*} \cos (\omega_1 t) \\
        0 && 2 g_s^{cv} \cos (\omega_1 t) && \omega_0 - i \Gamma
    \end{matrix} \right),
\end{gather}
where we make the rotational wave approximation for the cavity mode. To solve the Schrödinger equation 
\begin{gather}
i\frac{\partial \vec{\psi}}{\partial t} = \hat{H}_- \vec{\psi},
\end{gather}
we change variable $\vec{\psi} = ( \phi_1, \phi_2, \phi_3 \e^{i\omega_1 t})$ and get:
\begin{align*}
    i \dot{\phi}_1 &= (E_1-i\gamma_{\text{rad}}) \phi_1 + g_0^{vv} \phi_2, \\
    i \dot{\phi}_2 &= g_0^{vv*} \phi_1 + (E_{1'} + \omega_0 - i \gamma_{\text{rad}} - i \Gamma) \phi_2 + g_s^{cv*} (1 + \e^{i2\omega_1 t}) \phi_3, \\
    i \dot{\phi}_3 &= g_s^{cv} (\e^{-i2\omega_1 t}+ 1) \phi_2 + (\omega_0 + \omega_1 - i \Gamma) \phi_3.
\end{align*}
The fast oscillating terms $\e^{i2\omega_1 t}$ can be neglecting same as in the rotation wave approximation. The components $\phi_{1,3}$ have similar energy. The state $\phi_2$ can be exclude from this system if $E_{1'} - \omega_1 \gg g_0^{vv}, g_s^{cv}$ and we can write effective Hamiltonian in basis of exciton states and photon in cavity mode $(\hat{a}^\dag_{1-} |0\rangle, \hat{b}^\dag_{0-} |0\rangle)$:
\begin{align}
\label{HeffB}
    \hat{H}_{\text{eff}} = \left( \begin{matrix}
        E_1-i\gamma_{\text{rad}} && - \frac{g_0^{vv} g_s^{cv*}}{E_{1'} - \omega_1} \\
        - \frac{g_0^{vv*} g_s^{cv}}{E_{1'} - \omega_1} &&  \omega_0 + \omega_1 - \frac{|g_s^{cv}|^2}{E_{1'} - \omega_1} - i \Gamma
    \end{matrix} \right),
\end{align}
obtained to the leading order, assuming that $E_{1,1'},\omega_{0,1} \gg g_{0,s}, \Gamma, \gamma_{\text{rad}}$. This effective Hamiltonian describes an SDC that is similar to the James-Cummings model with a resonance condition:
\begin{gather}
    \Delta_{\text{XB}} = \omega_0 + \omega_1 - \frac{|g_s^{cv}|^2}{E_{1'} - \omega_1} - E_1,
\end{gather}
where $\Delta_{\text{XB}}$  is the detuning. 
The $\omega_0$ photon emission rate is:
\begin{gather}
\label{gammaSDCb}
    \gamma_{\text{SDC}}^{\text{XB}} = \frac{2 |g_0^{vv}|^2 |g_s^{cv}|^2 \Gamma }{\hbar (E_{1'} - \omega_1)^2 (\Gamma^2 + \Delta_{\text{XB}}^2)},
\end{gather}
where we also assume that $\gamma_{\text{rad}} \ll \Gamma$, otherwise we replace $\Gamma \rightarrow \Gamma - \gamma_{\text{rad}}$. The probability of emitting $\omega_0$ photon $\mathcal{P}_\text{BE}$ is the ratio between the recombination rate and all other state decay channels. For a bright exciton, the main decay process is radiative recombination:
\begin{gather}
    \mathcal{P}_\text{XB} = \frac{\gamma_{\text{SDC}}^{\text{XB}}}{\gamma_{\text{SDC}}^{\text{XB}} + \gamma_{\text{rad}}}.
\end{gather}

\begin{figure*}
\includegraphics[width=0.6\linewidth]{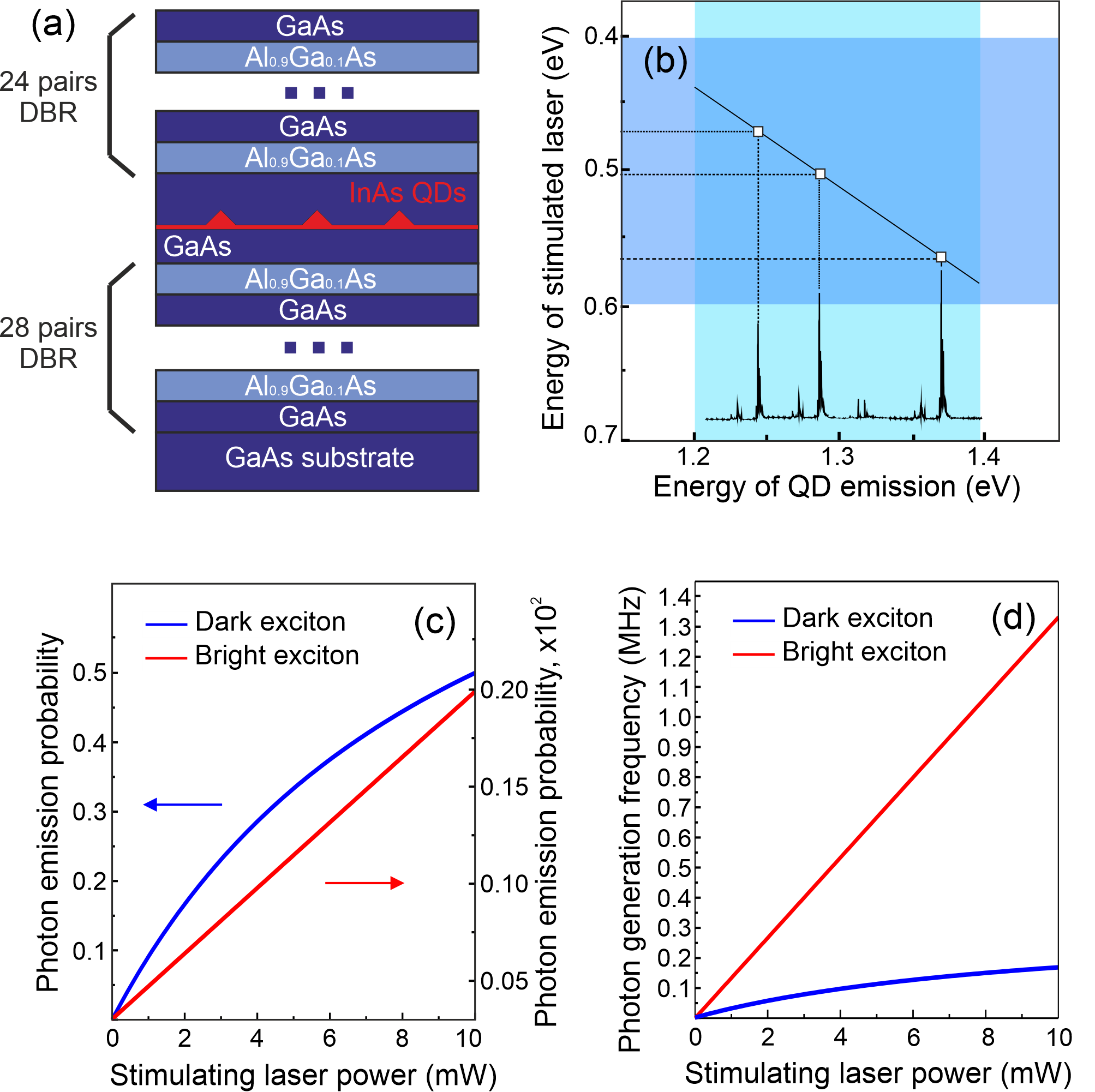}
\centering
\caption{\label{figStructure} (a) Schematic representation of a heterostructure suitable for forming a columnar microcavity with InAs QDs, where stimulated down-conversion for the C-band can be realized. (b) Narrow-line exciton spectrum typical for the emission range of InAs/GaAs QDs. The required tuning range of the stimulating laser energy is shown on the vertical axis. The dotted lines show the corresponding excitation laser energies for particular emission lines. (c) Power dependence of the emission probability for bright and dark excitons. (d) Power dependence of the SDC photon generation frequency for the bright and dark excitons.}
\end{figure*}

\textbf{Dark exciton.}
For the initial state of a dark exciton ($X_D$) given by equation (\ref{psiDE}), the stimulating beam must be circularly polarized. However, to obtain maximum efficiency, one can use linear polarization $\vect{e}_1=(\vect{e}_+ + \vect{e}_-)/\sqrt{2}$, which  can be presented by the sum of two sercularly polarized components of $\pm$  signs. The Hamiltonian for this case is:
\begin{gather} \nonumber
    \hat{H} = \hat{H}_+ + \hat{H}_-, \\ \nonumber
    \hat{H}_+ = (E_{1'} - i \gamma_{\text{rad}} )\hat{a}^\dag_{1'+} \hat{a}_{1'+} + 
    E_2 \hat{a}^\dag_{2+} \hat{a}_{2+} + 
    (\omega_0 - i \Gamma) \hat{b}^\dag_{0+} \hat{b}_{0+}  \\ \nonumber
    +(\hat{b}^\dag_{0+} + \hat{b}_{0+}) (g_0^{vv} \hat{a}^\dag_{1'+} \hat{a}_{2+} + g_0^{vv*} \hat{a}^\dag_{2+} \hat{a}_{1'+})  \\ \nonumber
    +2\cos (\omega_1 t) (g_s^{vv} \hat{a}^\dag_{1'+} \hat{a}_{2+} + g_s^{vv*} \hat{a}^\dag_{2+} \hat{a}_{1'+})  \\ \nonumber
    +(\hat{b}^\dag_{0+} + \hat{b}_{0+}) (g_0^{cv} \hat{a}^\dag_{1'+} + g_0^{cv*} \hat{a}_{1'+}) \\
    +2\cos (\omega_1 t) (g_s^{cv} \hat{a}^\dag_{1'+} + g_s^{cv*} \hat{a}_{1'+}),
\end{gather}
where $E_2$ is the dark exciton energy, $g_0^{cv(vv)}$ is the coupling constant between the mode and the exciton state, $g_s^{cv(vv)}$ is the interaction between the stimulating laser beam and the excitons state. 
By performing the same calculations as in the previous section in the basis of states $(\hat{a}^\dag_{2+} |0\rangle, \hat{a}^\dag_{1'+} |0\rangle, \hat{b}^\dag_{0+} \hat{a}^\dag_{1'+} |0\rangle, \hat{b}^\dag_{0+} |0\rangle)$, we can obtain the effective Hamiltonian for the states $(\hat{a}^\dag_{2+} |0\rangle, \hat{b}^\dag_{0+} |0\rangle)$:
\begin{align}
\label{HeffD}
    \hat{H}_{\text{eff}} = \left( \begin{matrix}
        E_2 - \frac{|g_s^{vv}|^2}{E_{1'} - \omega_0} && 
        - \frac{g_0^{vv} g_s^{cv*}}{E_{1'} - \omega_1} - \frac{g_s^{vv} g_0^{cv*}}{E_{1'} - \omega_0} \\
        - \frac{g_0^{vv*} g_s^{cv}}{E_{1'} - \omega_1} - \frac{g_s^{vv*} g_0^{cv}}{E_{1'} - \omega_0} &&  
        \omega_0 + \omega_1 - \frac{|g_s^{cv}|^2}{E_{1'} - \omega_1} - i \Gamma
    \end{matrix} \right).
\end{align}
The expressions for the detuning and the photon emission rate $\omega_0$ are:
\begin{gather}
    \Delta_{\text{XD}} = \omega_0 + \omega_1 - E_2 - \frac{|g_s^{cv}|^2}{E_{1'} - \omega_1}  + \frac{|g_s^{vv}|^2}{E_{1'} - \omega_0}, \\
    \gamma_{\text{SDC}}^{\text{XD}} = \frac{2\Gamma}{\hbar (\Gamma^2 + \Delta^2_{\text{XD}})} \left|
    \frac{g_0^{vv} g_s^{cv*}}{E_{1'} - \omega_1} + \frac{g_s^{vv} g_0^{cv*}}{E_{1'} - \omega_0} \right|^2.
\end{gather}
The emitted photon from the initial state (\ref{psiDE}) will be linear polarized with the polarization parallel to that of the stimulating beam due to different sign in $p_{cv}$ matrix elements and antisymmetric initial state. 
Same as for bright exciton we can introduce $\omega_0$ photon emission probability
\begin{gather}
    \mathcal{P}_\text{XD} = \frac{\gamma_{\text{SDC}}^{\text{XD}}}{\gamma_{\text{SDC}}^{\text{XD}} + \gamma_{\text{XD}}},
\end{gather}
 with the dark exciton lifetime $\tau_{\text{XD}} = 1/ \gamma_{\text{XD}}$ is determined by the non-SDC process.

\section{Discussion }

The single photon emission rate at $\omega_0$ strongly depends on the microresonator characteristics, which should have a high Q-factor to increase $\gamma_{\text{SDC}}$ by orders of magnitude. QD-embedded micropillar structures have been widely used for the C-band \cite{Musial2021, Wells2023}. With two highly reflective DBRs, they can achieve a Q-factor of $10^4 - 10^5$ \cite{Rakhlin2023, Schneider2016}. We performed numerical simulations to estimate the Q-factor at 1550 nm. The active region ($\lambda$-cavity) is assumed to be 470 nm thick and is surrounded by DBRs, which consist of 28 lower and 24 upper quarter-wave GaAs/Al$_{0.9}$Ga$_{0.1}$As layers (Fig. 3a). The Q-factor estimated from the resonance dip at 1550 nm in the calculated reflectance spectrum is 9300 for a planar heterostructure. For a zero-dimensional cavity in a micropillar, the value should increase.

An important advantage of our single-photon sources is the ability to obtain the required emission wavelength, in particular 1550 nm, despite the wide spread of QD emission energies. The energy distribution of InAs/GaAs QDs typically lies in the range of 1.2–1.4 eV \cite{Kitamura2015, Hu2020}. A measured QD emission spectrum is shown in Fig. 3b along with the required energy of the stimulating laser, which for this demonstrative spectrum should be tunable in the range of 0.4–0.6 eV. For example,  commercial lasers are available to obtain single-photon emission of 1.55 $\mu$m (0.8 eV) for any QD lines in the spectrum of Fig. 3b.

The efficiency of single photon generation in the proposed scheme is controlled by two factors: the emission probability and the photon generation rate in the SDC process. Both are different for bright and dark excitons and both depend on the excitation power of the stimulating laser. To estimate the single photon emission rate at 1.55 $\mu$m, according to equation~(\ref{gammaSDCb}), we used the following parameters: $g_0 = 20~\mu$eV, $\Gamma = 100~\mu$eV, $E_0 - \omega_1 = 0.8~$eV, $\varepsilon \approx 11$, $p_{cv}/mc = 10^{-2}$ \cite{Ivchenko_2014} and zero detuning. The stimulating laser can be focused on a square region $L^2 = (\lambda_1/2)^2$, and its power can be estimated as $P \approx S_p (\lambda_1/2)^2$. Thus, the emission rate of single photons is determined by the relation:
\begin{gather}
    \gamma_{\text{SDC}} \approx 10^8 P,
\end{gather}
where $\gamma_{\text{SDC}}$ is measured in 1/s and $P$ in watts (W).

The experimentally measured dark exciton lifetime for InAs/GaAs quantum dots is $\tau_{\text{DE}} \sim 1~\mu$s \cite{Schawrtz2015}. The radiative lifetime of bright excitons is estimated as $\tau_{\text{BE}} \sim 0.5 - 1$ ns \cite{Lee1998, Kono2005}. Using these parameters, the photon generation frequency can be calculated as:
\begin{gather}
   f_\text{SDC} = \frac{\mathcal{P}_\text{SDC}}{\tau_{\text{init}}},
\end{gather}
where $\tau_{\text{init}}$ is the time required to prepare the initial state. We set this value to $3 \cdot \tau_{\text{lifetime}}$, which corresponds to a probability greater than 95\% that there will be no charge carriers left in the QD. Accordingly, for a light exciton, the initialization time $\tau_{\text{init}}^{\text{BE}}$ is 1.5 ns, and for a dark exciton $\tau_{\text{init}}^{\text{DE}}$ is 3 $\mu$s.

The dependences of the photon emission probability and the single photon generation rate in the SDC process for bright and dark excitons are shown in Figure 3 (c,d). These dependences are plotted for a stimulating laser power of no more than 10 mW to avoid heating the sample. The probability of emitting a photon with an energy of $\hbar \omega_0$ for a dark exciton is two orders of magnitude higher than for a bright exciton, which is due to the longer lifetime of the dark exciton (Fig. 3c). In contrast, the photon generation rate for a bright exciton is almost an order of magnitude higher than for a dark exciton due to a faster initialization time $\tau_{\text{init}}$. The rate can reaches 1.3 MHz at a stimulating laser power of 10 mW (Fig. 3d). 

These evaluations demonstrate promising performance of single photon sources based on the SDC process in a quantum dot-cavity system.
As discussed in the introduction, alternative approaches to realize single-photon emission in the C-band cannot provide such amazing parameters. SPDC and SFWM sources are characterized by random generation times of photon pairs and low photon generation rates, which require expensive multiplexing \cite{Wang2024}.
Semiconductor QDs, such as InAs/InP or InAs/InGaAs QDs grown on a metamorphic buffer layer \cite{Holewa2020, Wyborski2023}, can provide sufficient purity and brightness of single photons. However, their practical implementation faces significant difficulties in the fabrication of resonator structures with QDs for the C-band. The frequency tunability of such QD sources is too small. Moreover, the raw two-photon interference visibility, which characterizes the indistinguishability of photons, does not exceed 20$\%$ due to the long lifetime ($\geq$ 1 ns) \cite{Holewa2024}. In our approach, the indistinguishability can reach much higher values because of the shorter lifetime of the bright exciton $\sim$ 500 ps and the coherence time $T_2 \sim$ 830 ps in the QDs emitting around 900 nm \cite{Galimov2021}.

\quad
\section{Conclusion}

Despite some uncertainty, which exists before practical implementation for any new approach, it can be concluded that the considered quantum down-conversion based on nonlinear interaction of quantum levels in a QD placed in a microcavity under a stimulating laser field is very promising for obtaining single photons of the required frequency on demand. The flexibility of the proposed approach is ensured by a special microcavity design for the target frequency of the fundamental optical mode and tuning the laser frequency to implement the difference frequency of single photon emission. This approach is very promising for providing identical sources of single photons, which is necessary for extended quantum key distribution systems or one-way quantum repeaters. High purity and indistinguishability of single photons are ensured by the ideal characteristics of a single QD fabricated using a well-established technology for 900 nm heterostructures. The output power during stimulated down-conversion significantly exceeds that in devices using spontaneous SPDC or SFWM processes. These are obvious advantages rarely realized for the telecommunication range. Further development of our method for charged QDs may prove promising for the formation of multiphoton cluster states, which are so necessary for efficient photonic quantum computing. Ultimately, the proposed scheme provides an efficient interface between ideal QD-based emitters and a long-range telecommunication network that extends the capabilities of single-photon sources to advanced quantum-photonic functions.

\section{ACKNOWLEDGMENTS}

I.V.K. acknowledge the partial support of the Russian Science Foundation (analytical theory–Project No. 22-12-00139). We also thank N.S.~Averkiev for fruitful discussions.

\bibliography{bibl}

\begin{thebibliography}{42}%
\makeatletter
\providecommand \@ifxundefined [1]{%
 \@ifx{#1\undefined}
}%
\providecommand \@ifnum [1]{%
 \ifnum #1\expandafter \@firstoftwo
 \else \expandafter \@secondoftwo
 \fi
}%
\providecommand \@ifx [1]{%
 \ifx #1\expandafter \@firstoftwo
 \else \expandafter \@secondoftwo
 \fi
}%
\providecommand \natexlab [1]{#1}%
\providecommand \enquote  [1]{``#1''}%
\providecommand \bibnamefont  [1]{#1}%
\providecommand \bibfnamefont [1]{#1}%
\providecommand \citenamefont [1]{#1}%
\providecommand \href@noop [0]{\@secondoftwo}%
\providecommand \href [0]{\begingroup \@sanitize@url \@href}%
\providecommand \@href[1]{\@@startlink{#1}\@@href}%
\providecommand \@@href[1]{\endgroup#1\@@endlink}%
\providecommand \@sanitize@url [0]{\catcode `\\12\catcode `\$12\catcode
  `\&12\catcode `\#12\catcode `\^12\catcode `\_12\catcode `\%12\relax}%
\providecommand \@@startlink[1]{}%
\providecommand \@@endlink[0]{}%
\providecommand \url  [0]{\begingroup\@sanitize@url \@url }%
\providecommand \@url [1]{\endgroup\@href {#1}{\urlprefix }}%
\providecommand \urlprefix  [0]{URL }%
\providecommand \Eprint [0]{\href }%
\providecommand \doibase [0]{https://doi.org/}%
\providecommand \selectlanguage [0]{\@gobble}%
\providecommand \bibinfo  [0]{\@secondoftwo}%
\providecommand \bibfield  [0]{\@secondoftwo}%
\providecommand \translation [1]{[#1]}%
\providecommand \BibitemOpen [0]{}%
\providecommand \bibitemStop [0]{}%
\providecommand \bibitemNoStop [0]{.\EOS\space}%
\providecommand \EOS [0]{\spacefactor3000\relax}%
\providecommand \BibitemShut  [1]{\csname bibitem#1\endcsname}%
\let\auto@bib@innerbib\@empty
\bibitem [{\citenamefont {Senellart}\ \emph {et~al.}(2017)\citenamefont
  {Senellart}, \citenamefont {Solomon},\ and\ \citenamefont
  {White}}]{Senellart2017}%
  \BibitemOpen
  \bibfield  {author} {\bibinfo {author} {\bibfnamefont {P.}~\bibnamefont
  {Senellart}}, \bibinfo {author} {\bibfnamefont {G.}~\bibnamefont {Solomon}},\
  and\ \bibinfo {author} {\bibfnamefont {A.}~\bibnamefont {White}},\ }\bibfield
   {title} {\bibinfo {title} {High-performance semiconductor quantum-dot
  single-photon sources},\ }\href@noop {} {\bibfield  {journal} {\bibinfo
  {journal} {Nature Nanotechnology}\ }\textbf {\bibinfo {volume} {12}},\
  \bibinfo {pages} {1026} (\bibinfo {year} {2017})}\BibitemShut {NoStop}%
\bibitem [{\citenamefont {Rakhlin}\ \emph {et~al.}(2023)\citenamefont
  {Rakhlin}, \citenamefont {Galimov}, \citenamefont {Dyakonov}, \citenamefont
  {Skryabin}, \citenamefont {Klimko}, \citenamefont {Kulagina}, \citenamefont
  {Zadiranov}, \citenamefont {Sorokin}, \citenamefont {Sedova}, \citenamefont
  {Guseva}, \citenamefont {Berezina}, \citenamefont {Serov}, \citenamefont
  {Maleev}, \citenamefont {Kuzmenkov}, \citenamefont {Troshkov}, \citenamefont
  {Taratorin}, \citenamefont {Skalkin}, \citenamefont {Straupe}, \citenamefont
  {Kulik}, \citenamefont {Shubina},\ and\ \citenamefont
  {Toropov}}]{Rakhlin2023}%
  \BibitemOpen
  \bibfield  {author} {\bibinfo {author} {\bibfnamefont {M.}~\bibnamefont
  {Rakhlin}}, \bibinfo {author} {\bibfnamefont {A.}~\bibnamefont {Galimov}},
  \bibinfo {author} {\bibfnamefont {I.}~\bibnamefont {Dyakonov}}, \bibinfo
  {author} {\bibfnamefont {N.}~\bibnamefont {Skryabin}}, \bibinfo {author}
  {\bibfnamefont {G.}~\bibnamefont {Klimko}}, \bibinfo {author} {\bibfnamefont
  {M.}~\bibnamefont {Kulagina}}, \bibinfo {author} {\bibfnamefont
  {Y.}~\bibnamefont {Zadiranov}}, \bibinfo {author} {\bibfnamefont
  {S.}~\bibnamefont {Sorokin}}, \bibinfo {author} {\bibfnamefont
  {I.}~\bibnamefont {Sedova}}, \bibinfo {author} {\bibfnamefont
  {Y.}~\bibnamefont {Guseva}}, \bibinfo {author} {\bibfnamefont
  {D.}~\bibnamefont {Berezina}}, \bibinfo {author} {\bibfnamefont
  {Y.}~\bibnamefont {Serov}}, \bibinfo {author} {\bibfnamefont
  {N.}~\bibnamefont {Maleev}}, \bibinfo {author} {\bibfnamefont
  {A.}~\bibnamefont {Kuzmenkov}}, \bibinfo {author} {\bibfnamefont
  {S.}~\bibnamefont {Troshkov}}, \bibinfo {author} {\bibfnamefont
  {K.}~\bibnamefont {Taratorin}}, \bibinfo {author} {\bibfnamefont
  {A.}~\bibnamefont {Skalkin}}, \bibinfo {author} {\bibfnamefont
  {S.}~\bibnamefont {Straupe}}, \bibinfo {author} {\bibfnamefont
  {S.}~\bibnamefont {Kulik}}, \bibinfo {author} {\bibfnamefont
  {T.}~\bibnamefont {Shubina}},\ and\ \bibinfo {author} {\bibfnamefont
  {A.}~\bibnamefont {Toropov}},\ }\bibfield  {title} {\bibinfo {title}
  {Demultiplexed single-photon source with a quantum dot coupled to
  microresonator},\ }\href@noop {} {\bibfield  {journal} {\bibinfo  {journal}
  {Journal of Luminescence}\ }\textbf {\bibinfo {volume} {253}},\ \bibinfo
  {pages} {119496} (\bibinfo {year} {2023})}\BibitemShut {NoStop}%
\bibitem [{\citenamefont {Vajner}\ \emph {et~al.}(2022)\citenamefont {Vajner},
  \citenamefont {Rickert}, \citenamefont {Gao}, \citenamefont {Kaymazlar},\
  and\ \citenamefont {Heindel}}]{Vajner2022}%
  \BibitemOpen
  \bibfield  {author} {\bibinfo {author} {\bibfnamefont {D.}~\bibnamefont
  {Vajner}}, \bibinfo {author} {\bibfnamefont {L.}~\bibnamefont {Rickert}},
  \bibinfo {author} {\bibfnamefont {T.}~\bibnamefont {Gao}}, \bibinfo {author}
  {\bibfnamefont {K.}~\bibnamefont {Kaymazlar}},\ and\ \bibinfo {author}
  {\bibfnamefont {T.}~\bibnamefont {Heindel}},\ }\bibfield  {title} {\bibinfo
  {title} {Quantum communication using semiconductor quantum dots},\
  }\href@noop {} {\bibfield  {journal} {\bibinfo  {journal} {Advanced Quantum
  Technologies}\ }\textbf {\bibinfo {volume} {5}},\ \bibinfo {pages} {2100116}
  (\bibinfo {year} {2022})}\BibitemShut {NoStop}%
\bibitem [{\citenamefont {Caspani}\ \emph {et~al.}(2017)\citenamefont
  {Caspani}, \citenamefont {Xiong}, \citenamefont {Eggleton}, \citenamefont
  {Bajoni}, \citenamefont {Liscidini}, \citenamefont {Galli}, \citenamefont
  {Morandotti},\ and\ \citenamefont {Moss}}]{Caspani2017}%
  \BibitemOpen
  \bibfield  {author} {\bibinfo {author} {\bibfnamefont {L.}~\bibnamefont
  {Caspani}}, \bibinfo {author} {\bibfnamefont {C.}~\bibnamefont {Xiong}},
  \bibinfo {author} {\bibfnamefont {B.}~\bibnamefont {Eggleton}}, \bibinfo
  {author} {\bibfnamefont {D.}~\bibnamefont {Bajoni}}, \bibinfo {author}
  {\bibfnamefont {M.}~\bibnamefont {Liscidini}}, \bibinfo {author}
  {\bibfnamefont {M.}~\bibnamefont {Galli}}, \bibinfo {author} {\bibfnamefont
  {R.}~\bibnamefont {Morandotti}},\ and\ \bibinfo {author} {\bibfnamefont
  {D.}~\bibnamefont {Moss}},\ }\bibfield  {title} {\bibinfo {title} {Integrated
  sources of photon quantum states based on nonlinear optics},\ }\href@noop {}
  {\bibfield  {journal} {\bibinfo  {journal} {Light: Science and Applications}\
  }\textbf {\bibinfo {volume} {6}},\ \bibinfo {pages} {e17100} (\bibinfo {year}
  {2017})}\BibitemShut {NoStop}%
\bibitem [{\citenamefont {Baboux}\ \emph {et~al.}(2023)\citenamefont {Baboux},
  \citenamefont {Moody},\ and\ \citenamefont {Ducci}}]{Baboux2023}%
  \BibitemOpen
  \bibfield  {author} {\bibinfo {author} {\bibfnamefont {F.}~\bibnamefont
  {Baboux}}, \bibinfo {author} {\bibfnamefont {G.}~\bibnamefont {Moody}},\ and\
  \bibinfo {author} {\bibfnamefont {S.}~\bibnamefont {Ducci}},\ }\bibfield
  {title} {\bibinfo {title} {Nonlinear integrated quantum photonics with
  {A}l{G}a{A}s},\ }\href@noop {} {\bibfield  {journal} {\bibinfo  {journal}
  {Optica}\ }\textbf {\bibinfo {volume} {10}},\ \bibinfo {pages} {917}
  (\bibinfo {year} {2023})}\BibitemShut {NoStop}%
\bibitem [{\citenamefont {Wang}\ \emph {et~al.}(2024)\citenamefont {Wang},
  \citenamefont {Zeng}, \citenamefont {Ma},\ and\ \citenamefont
  {Yuan}}]{Wang2024}%
  \BibitemOpen
  \bibfield  {author} {\bibinfo {author} {\bibfnamefont {H.}~\bibnamefont
  {Wang}}, \bibinfo {author} {\bibfnamefont {Q.}~\bibnamefont {Zeng}}, \bibinfo
  {author} {\bibfnamefont {H.}~\bibnamefont {Ma}},\ and\ \bibinfo {author}
  {\bibfnamefont {Z.}~\bibnamefont {Yuan}},\ }\bibfield  {title} {\bibinfo
  {title} {Progress on chip-based spontaneous four-wave mixing quantum light
  sources},\ }\href@noop {} {\bibfield  {journal} {\bibinfo  {journal}
  {Advanced Devices and Instrumentation}\ }\textbf {\bibinfo {volume} {5}},\
  \bibinfo {pages} {0032} (\bibinfo {year} {2024})}\BibitemShut {NoStop}%
\bibitem [{\citenamefont {Wang}\ \emph {et~al.}(2019)\citenamefont {Wang},
  \citenamefont {He}, \citenamefont {Chung}, \citenamefont {Hu}, \citenamefont
  {Yu}, \citenamefont {Chen}, \citenamefont {Ding}, \citenamefont {Chen},
  \citenamefont {Qin}, \citenamefont {Yang}, \citenamefont {Liu}, \citenamefont
  {Duan}, \citenamefont {Li}, \citenamefont {Gerhardt}, \citenamefont
  {Winkler}, \citenamefont {Jurkat}, \citenamefont {Wang}, \citenamefont
  {Gregersen}, \citenamefont {Huo}, \citenamefont {Dai}, \citenamefont {Yu},
  \citenamefont {H\"ofling}, \citenamefont {Lu},\ and\ \citenamefont
  {Pan}}]{Wang2019}%
  \BibitemOpen
  \bibfield  {author} {\bibinfo {author} {\bibfnamefont {H.}~\bibnamefont
  {Wang}}, \bibinfo {author} {\bibfnamefont {Y.-M.}\ \bibnamefont {He}},
  \bibinfo {author} {\bibfnamefont {T.-H.}\ \bibnamefont {Chung}}, \bibinfo
  {author} {\bibfnamefont {H.}~\bibnamefont {Hu}}, \bibinfo {author}
  {\bibfnamefont {Y.}~\bibnamefont {Yu}}, \bibinfo {author} {\bibfnamefont
  {S.}~\bibnamefont {Chen}}, \bibinfo {author} {\bibfnamefont {X.}~\bibnamefont
  {Ding}}, \bibinfo {author} {\bibfnamefont {M.-C.}\ \bibnamefont {Chen}},
  \bibinfo {author} {\bibfnamefont {J.}~\bibnamefont {Qin}}, \bibinfo {author}
  {\bibfnamefont {X.}~\bibnamefont {Yang}}, \bibinfo {author} {\bibfnamefont
  {R.-Z.}\ \bibnamefont {Liu}}, \bibinfo {author} {\bibfnamefont {Z.-C.}\
  \bibnamefont {Duan}}, \bibinfo {author} {\bibfnamefont {J.-P.}\ \bibnamefont
  {Li}}, \bibinfo {author} {\bibfnamefont {S.}~\bibnamefont {Gerhardt}},
  \bibinfo {author} {\bibfnamefont {K.}~\bibnamefont {Winkler}}, \bibinfo
  {author} {\bibfnamefont {J.}~\bibnamefont {Jurkat}}, \bibinfo {author}
  {\bibfnamefont {L.-J.}\ \bibnamefont {Wang}}, \bibinfo {author}
  {\bibfnamefont {N.}~\bibnamefont {Gregersen}}, \bibinfo {author}
  {\bibfnamefont {Y.-H.}\ \bibnamefont {Huo}}, \bibinfo {author} {\bibfnamefont
  {Q.}~\bibnamefont {Dai}}, \bibinfo {author} {\bibfnamefont {S.}~\bibnamefont
  {Yu}}, \bibinfo {author} {\bibfnamefont {S.}~\bibnamefont {H\"ofling}},
  \bibinfo {author} {\bibfnamefont {C.-Y.}\ \bibnamefont {Lu}},\ and\ \bibinfo
  {author} {\bibfnamefont {J.-W.}\ \bibnamefont {Pan}},\ }\bibfield  {title}
  {\bibinfo {title} {Towards optimal single-photon sources from polarized
  microcavities},\ }\href@noop {} {\bibfield  {journal} {\bibinfo  {journal}
  {Nature Photonics}\ }\textbf {\bibinfo {volume} {13}},\ \bibinfo {pages}
  {770–775} (\bibinfo {year} {2019})}\BibitemShut {NoStop}%
\bibitem [{\citenamefont {Tomm}\ \emph {et~al.}(2021)\citenamefont {Tomm},
  \citenamefont {Javadi}, \citenamefont {Antoniadis}, \citenamefont {Najer},
  \citenamefont {L\"obl}, \citenamefont {Korsch}, \citenamefont {Schott},
  \citenamefont {Valentin}, \citenamefont {Wieck}, \citenamefont {Ludwig},\
  and\ \citenamefont {Warburton}}]{Tomm2021}%
  \BibitemOpen
  \bibfield  {author} {\bibinfo {author} {\bibfnamefont {N.}~\bibnamefont
  {Tomm}}, \bibinfo {author} {\bibfnamefont {A.}~\bibnamefont {Javadi}},
  \bibinfo {author} {\bibfnamefont {N.}~\bibnamefont {Antoniadis}}, \bibinfo
  {author} {\bibfnamefont {D.}~\bibnamefont {Najer}}, \bibinfo {author}
  {\bibfnamefont {M.}~\bibnamefont {L\"obl}}, \bibinfo {author} {\bibfnamefont
  {A.}~\bibnamefont {Korsch}}, \bibinfo {author} {\bibfnamefont
  {R.}~\bibnamefont {Schott}}, \bibinfo {author} {\bibfnamefont
  {S.}~\bibnamefont {Valentin}}, \bibinfo {author} {\bibfnamefont
  {A.}~\bibnamefont {Wieck}}, \bibinfo {author} {\bibfnamefont
  {A.}~\bibnamefont {Ludwig}},\ and\ \bibinfo {author} {\bibfnamefont
  {R.}~\bibnamefont {Warburton}},\ }\bibfield  {title} {\bibinfo {title} {A
  bright and fast source of coherent single photons},\ }\href@noop {}
  {\bibfield  {journal} {\bibinfo  {journal} {Nature Nanotechnology}\ }\textbf
  {\bibinfo {volume} {16}},\ \bibinfo {pages} {399} (\bibinfo {year}
  {2021})}\BibitemShut {NoStop}%
\bibitem [{\citenamefont {Ding}\ \emph {et~al.}(2023)\citenamefont {Ding},
  \citenamefont {Guo}, \citenamefont {Xu}, \citenamefont {Liu}, \citenamefont
  {Zou}, \citenamefont {Zhao}, \citenamefont {Ge}, \citenamefont {Zhang},
  \citenamefont {Liu}, \citenamefont {Wang}, \citenamefont {Chen},
  \citenamefont {Wang}, \citenamefont {He}, \citenamefont {Huo}, \citenamefont
  {Lu},\ and\ \citenamefont {Pan}}]{Ding2023}%
  \BibitemOpen
  \bibfield  {author} {\bibinfo {author} {\bibfnamefont {X.}~\bibnamefont
  {Ding}}, \bibinfo {author} {\bibfnamefont {Y.-P.}\ \bibnamefont {Guo}},
  \bibinfo {author} {\bibfnamefont {M.-C.}\ \bibnamefont {Xu}}, \bibinfo
  {author} {\bibfnamefont {R.-Z.}\ \bibnamefont {Liu}}, \bibinfo {author}
  {\bibfnamefont {G.-Y.}\ \bibnamefont {Zou}}, \bibinfo {author} {\bibfnamefont
  {J.-Y.}\ \bibnamefont {Zhao}}, \bibinfo {author} {\bibfnamefont {Z.-X.}\
  \bibnamefont {Ge}}, \bibinfo {author} {\bibfnamefont {Q.-H.}\ \bibnamefont
  {Zhang}}, \bibinfo {author} {\bibfnamefont {H.-L.}\ \bibnamefont {Liu}},
  \bibinfo {author} {\bibfnamefont {L.-J.}\ \bibnamefont {Wang}}, \bibinfo
  {author} {\bibfnamefont {M.-C.}\ \bibnamefont {Chen}}, \bibinfo {author}
  {\bibfnamefont {H.}~\bibnamefont {Wang}}, \bibinfo {author} {\bibfnamefont
  {Y.-M.}\ \bibnamefont {He}}, \bibinfo {author} {\bibfnamefont {Y.-H.}\
  \bibnamefont {Huo}}, \bibinfo {author} {\bibfnamefont {C.-Y.}\ \bibnamefont
  {Lu}},\ and\ \bibinfo {author} {\bibfnamefont {J.-W.}\ \bibnamefont {Pan}},\
  }\href@noop {} {\bibinfo {title} {High-efficiency single-photon source above
  the loss-tolerant threshold for efficient linear optical quantum computing}}
  (\bibinfo {year} {2023}),\ \Eprint {https://arxiv.org/abs/2311.08347}
  {arXiv:2311.08347} \BibitemShut {NoStop}%
\bibitem [{\citenamefont {Nawrath}\ \emph {et~al.}(2023)\citenamefont
  {Nawrath}, \citenamefont {Joos}, \citenamefont {Kolatschek}, \citenamefont
  {Bauer}, \citenamefont {Pruy}, \citenamefont {Hornung}, \citenamefont
  {Fischer}, \citenamefont {Huang}, \citenamefont {Vijayan}, \citenamefont
  {Sittig}, \citenamefont {Jetter}, \citenamefont {Portalupi},\ and\
  \citenamefont {Michler}}]{Nawrath2023}%
  \BibitemOpen
  \bibfield  {author} {\bibinfo {author} {\bibfnamefont {C.}~\bibnamefont
  {Nawrath}}, \bibinfo {author} {\bibfnamefont {R.}~\bibnamefont {Joos}},
  \bibinfo {author} {\bibfnamefont {S.}~\bibnamefont {Kolatschek}}, \bibinfo
  {author} {\bibfnamefont {S.}~\bibnamefont {Bauer}}, \bibinfo {author}
  {\bibfnamefont {P.}~\bibnamefont {Pruy}}, \bibinfo {author} {\bibfnamefont
  {F.}~\bibnamefont {Hornung}}, \bibinfo {author} {\bibfnamefont
  {J.}~\bibnamefont {Fischer}}, \bibinfo {author} {\bibfnamefont
  {J.}~\bibnamefont {Huang}}, \bibinfo {author} {\bibfnamefont
  {P.}~\bibnamefont {Vijayan}}, \bibinfo {author} {\bibfnamefont
  {R.}~\bibnamefont {Sittig}}, \bibinfo {author} {\bibfnamefont
  {M.}~\bibnamefont {Jetter}}, \bibinfo {author} {\bibfnamefont
  {S.}~\bibnamefont {Portalupi}},\ and\ \bibinfo {author} {\bibfnamefont
  {P.}~\bibnamefont {Michler}},\ }\bibfield  {title} {\bibinfo {title} {Bright
  source of purcell-enhanced, triggered, single photons in the telecom
  {C}-band},\ }\href@noop {} {\bibfield  {journal} {\bibinfo  {journal}
  {Advanced Quantum Technologies}\ }\textbf {\bibinfo {volume} {6}},\ \bibinfo
  {pages} {2300111} (\bibinfo {year} {2023})}\BibitemShut {NoStop}%
\bibitem [{\citenamefont {Yang}\ \emph {et~al.}(2024)\citenamefont {Yang},
  \citenamefont {Chen}, \citenamefont {Rao}, \citenamefont {Zheng},
  \citenamefont {Song}, \citenamefont {Chen}, \citenamefont {Xiong},
  \citenamefont {Chen}, \citenamefont {Zhang}, \citenamefont {Wu},
  \citenamefont {Yu},\ and\ \citenamefont {Yu}}]{Yang2024}%
  \BibitemOpen
  \bibfield  {author} {\bibinfo {author} {\bibfnamefont {J.}~\bibnamefont
  {Yang}}, \bibinfo {author} {\bibfnamefont {Y.}~\bibnamefont {Chen}}, \bibinfo
  {author} {\bibfnamefont {Z.}~\bibnamefont {Rao}}, \bibinfo {author}
  {\bibfnamefont {Z.}~\bibnamefont {Zheng}}, \bibinfo {author} {\bibfnamefont
  {C.}~\bibnamefont {Song}}, \bibinfo {author} {\bibfnamefont {Y.}~\bibnamefont
  {Chen}}, \bibinfo {author} {\bibfnamefont {K.}~\bibnamefont {Xiong}},
  \bibinfo {author} {\bibfnamefont {P.}~\bibnamefont {Chen}}, \bibinfo {author}
  {\bibfnamefont {C.}~\bibnamefont {Zhang}}, \bibinfo {author} {\bibfnamefont
  {W.}~\bibnamefont {Wu}}, \bibinfo {author} {\bibfnamefont {Y.}~\bibnamefont
  {Yu}},\ and\ \bibinfo {author} {\bibfnamefont {S.}~\bibnamefont {Yu}},\
  }\bibfield  {title} {\bibinfo {title} {Tunable quantum dots in monolithic
  fabry-perot microcavities for high-performance single-photon sources},\
  }\href@noop {} {\bibfield  {journal} {\bibinfo  {journal} {Light: Science and
  Applications}\ }\textbf {\bibinfo {volume} {13}},\ \bibinfo {pages} {33}
  (\bibinfo {year} {2024})}\BibitemShut {NoStop}%
\bibitem [{\citenamefont {Paesani}\ \emph {et~al.}(2020)\citenamefont
  {Paesani}, \citenamefont {Borghi}, \citenamefont {Signorini}, \citenamefont
  {Mainos}, \citenamefont {Pavesi},\ and\ \citenamefont {Laing}}]{Paesani2020}%
  \BibitemOpen
  \bibfield  {author} {\bibinfo {author} {\bibfnamefont {S.}~\bibnamefont
  {Paesani}}, \bibinfo {author} {\bibfnamefont {M.}~\bibnamefont {Borghi}},
  \bibinfo {author} {\bibfnamefont {S.}~\bibnamefont {Signorini}}, \bibinfo
  {author} {\bibfnamefont {A.}~\bibnamefont {Mainos}}, \bibinfo {author}
  {\bibfnamefont {L.}~\bibnamefont {Pavesi}},\ and\ \bibinfo {author}
  {\bibfnamefont {A.}~\bibnamefont {Laing}},\ }\bibfield  {title} {\bibinfo
  {title} {Near-ideal spontaneous photon sources in silicon quantum
  photonics},\ }\href@noop {} {\bibfield  {journal} {\bibinfo  {journal}
  {Nature Communications}\ }\textbf {\bibinfo {volume} {11}},\ \bibinfo {pages}
  {2505} (\bibinfo {year} {2020})}\BibitemShut {NoStop}%
\bibitem [{\citenamefont {Helmy}\ \emph {et~al.}(2011)\citenamefont {Helmy},
  \citenamefont {Abolghasem}, \citenamefont {Aitchison}, \citenamefont
  {Bijlani}, \citenamefont {Han}, \citenamefont {Holmes}, \citenamefont
  {Hutchings}, \citenamefont {Younis},\ and\ \citenamefont
  {Wagner}}]{Helmy2011}%
  \BibitemOpen
  \bibfield  {author} {\bibinfo {author} {\bibfnamefont {A.}~\bibnamefont
  {Helmy}}, \bibinfo {author} {\bibfnamefont {P.}~\bibnamefont {Abolghasem}},
  \bibinfo {author} {\bibfnamefont {J.}~\bibnamefont {Aitchison}}, \bibinfo
  {author} {\bibfnamefont {B.}~\bibnamefont {Bijlani}}, \bibinfo {author}
  {\bibfnamefont {J.}~\bibnamefont {Han}}, \bibinfo {author} {\bibfnamefont
  {B.}~\bibnamefont {Holmes}}, \bibinfo {author} {\bibfnamefont
  {D.}~\bibnamefont {Hutchings}}, \bibinfo {author} {\bibfnamefont
  {U.}~\bibnamefont {Younis}},\ and\ \bibinfo {author} {\bibfnamefont
  {S.}~\bibnamefont {Wagner}},\ }\bibfield  {title} {\bibinfo {title} {Recent
  advances in phase matching of second-order nonlinearities in monolithic
  semiconductor waveguides},\ }\href@noop {} {\bibfield  {journal} {\bibinfo
  {journal} {Laser and Photonics Reviews}\ }\textbf {\bibinfo {volume} {5}},\
  \bibinfo {pages} {272–286} (\bibinfo {year} {2011})}\BibitemShut {NoStop}%
\bibitem [{\citenamefont {Ma}\ \emph {et~al.}(2011)\citenamefont {Ma},
  \citenamefont {Zotter}, \citenamefont {Kofler}, \citenamefont {Jennewein},\
  and\ \citenamefont {Zeilinger}}]{Ma2011}%
  \BibitemOpen
  \bibfield  {author} {\bibinfo {author} {\bibfnamefont {X.-s.}\ \bibnamefont
  {Ma}}, \bibinfo {author} {\bibfnamefont {S.}~\bibnamefont {Zotter}}, \bibinfo
  {author} {\bibfnamefont {J.}~\bibnamefont {Kofler}}, \bibinfo {author}
  {\bibfnamefont {T.}~\bibnamefont {Jennewein}},\ and\ \bibinfo {author}
  {\bibfnamefont {A.}~\bibnamefont {Zeilinger}},\ }\bibfield  {title} {\bibinfo
  {title} {Experimental generation of single photons via active multiplexing},\
  }\href@noop {} {\bibfield  {journal} {\bibinfo  {journal} {Physical Review
  A}\ }\textbf {\bibinfo {volume} {83}},\ \bibinfo {pages} {043814} (\bibinfo
  {year} {2011})}\BibitemShut {NoStop}%
\bibitem [{\citenamefont {Meyer-Scott}\ \emph {et~al.}(2020)\citenamefont
  {Meyer-Scott}, \citenamefont {Silberhorn},\ and\ \citenamefont
  {Migdall}}]{Scott2020}%
  \BibitemOpen
  \bibfield  {author} {\bibinfo {author} {\bibfnamefont {E.}~\bibnamefont
  {Meyer-Scott}}, \bibinfo {author} {\bibfnamefont {C.}~\bibnamefont
  {Silberhorn}},\ and\ \bibinfo {author} {\bibfnamefont {A.}~\bibnamefont
  {Migdall}},\ }\bibfield  {title} {\bibinfo {title} {Single-photon sources:
  {A}pproaching the ideal through multiplexing},\ }\href@noop {} {\bibfield
  {journal} {\bibinfo  {journal} {Review of Scientific Instruments}\ }\textbf
  {\bibinfo {volume} {91}},\ \bibinfo {pages} {041101} (\bibinfo {year}
  {2020})}\BibitemShut {NoStop}%
\bibitem [{\citenamefont {Spring}\ \emph {et~al.}(2013)\citenamefont {Spring},
  \citenamefont {Salter}, \citenamefont {Metcalf}, \citenamefont {Humphreys},
  \citenamefont {Moore}, \citenamefont {Thomas-Peter}, \citenamefont
  {Barbieri}, \citenamefont {Jin}, \citenamefont {Langford}, \citenamefont
  {Kolthammer}, \citenamefont {Booth},\ and\ \citenamefont
  {Walmsley}}]{Spring2013}%
  \BibitemOpen
  \bibfield  {author} {\bibinfo {author} {\bibfnamefont {J.}~\bibnamefont
  {Spring}}, \bibinfo {author} {\bibfnamefont {P.}~\bibnamefont {Salter}},
  \bibinfo {author} {\bibfnamefont {B.}~\bibnamefont {Metcalf}}, \bibinfo
  {author} {\bibfnamefont {P.}~\bibnamefont {Humphreys}}, \bibinfo {author}
  {\bibfnamefont {M.}~\bibnamefont {Moore}}, \bibinfo {author} {\bibfnamefont
  {N.}~\bibnamefont {Thomas-Peter}}, \bibinfo {author} {\bibfnamefont
  {M.}~\bibnamefont {Barbieri}}, \bibinfo {author} {\bibfnamefont {X.-M.}\
  \bibnamefont {Jin}}, \bibinfo {author} {\bibfnamefont {N.}~\bibnamefont
  {Langford}}, \bibinfo {author} {\bibfnamefont {W.}~\bibnamefont
  {Kolthammer}}, \bibinfo {author} {\bibfnamefont {M.}~\bibnamefont {Booth}},\
  and\ \bibinfo {author} {\bibfnamefont {I.}~\bibnamefont {Walmsley}},\
  }\bibfield  {title} {\bibinfo {title} {On-chip low loss heralded source of
  pure single photons},\ }\href@noop {} {\bibfield  {journal} {\bibinfo
  {journal} {Optics Express}\ }\textbf {\bibinfo {volume} {21}},\ \bibinfo
  {pages} {13522} (\bibinfo {year} {2013})}\BibitemShut {NoStop}%
\bibitem [{\citenamefont {Wang}\ \emph {et~al.}(2021)\citenamefont {Wang},
  \citenamefont {Jöns},\ and\ \citenamefont {Sun}}]{Wang2021}%
  \BibitemOpen
  \bibfield  {author} {\bibinfo {author} {\bibfnamefont {Y.}~\bibnamefont
  {Wang}}, \bibinfo {author} {\bibfnamefont {K.~D.}\ \bibnamefont {Jöns}},\
  and\ \bibinfo {author} {\bibfnamefont {Z.}~\bibnamefont {Sun}},\ }\bibfield
  {title} {\bibinfo {title} {Integrated photon-pair sources with nonlinear
  optics},\ }\href@noop {} {\bibfield  {journal} {\bibinfo  {journal} {Applied
  Physics Reviews}\ }\textbf {\bibinfo {volume} {8}},\ \bibinfo {pages}
  {011314} (\bibinfo {year} {2021})}\BibitemShut {NoStop}%
\bibitem [{\citenamefont {Eltes}\ \emph {et~al.}(2020)\citenamefont {Eltes},
  \citenamefont {Villarreal-Garcia}, \citenamefont {Caimi}, \citenamefont
  {Siegwart}, \citenamefont {Gentile}, \citenamefont {Hart}, \citenamefont
  {Stark}, \citenamefont {Marshall}, \citenamefont {Thompson}, \citenamefont
  {Barreto}, \citenamefont {Fompeyrine},\ and\ \citenamefont
  {Abel}}]{Eltes2020}%
  \BibitemOpen
  \bibfield  {author} {\bibinfo {author} {\bibfnamefont {F.}~\bibnamefont
  {Eltes}}, \bibinfo {author} {\bibfnamefont {G.}~\bibnamefont
  {Villarreal-Garcia}}, \bibinfo {author} {\bibfnamefont {D.}~\bibnamefont
  {Caimi}}, \bibinfo {author} {\bibfnamefont {H.}~\bibnamefont {Siegwart}},
  \bibinfo {author} {\bibfnamefont {A.}~\bibnamefont {Gentile}}, \bibinfo
  {author} {\bibfnamefont {A.}~\bibnamefont {Hart}}, \bibinfo {author}
  {\bibfnamefont {P.}~\bibnamefont {Stark}}, \bibinfo {author} {\bibfnamefont
  {G.}~\bibnamefont {Marshall}}, \bibinfo {author} {\bibfnamefont
  {M.}~\bibnamefont {Thompson}}, \bibinfo {author} {\bibfnamefont
  {J.}~\bibnamefont {Barreto}}, \bibinfo {author} {\bibfnamefont
  {J.}~\bibnamefont {Fompeyrine}},\ and\ \bibinfo {author} {\bibfnamefont
  {S.}~\bibnamefont {Abel}},\ }\bibfield  {title} {\bibinfo {title} {An
  integrated optical modulator operating at cryogenic temperatures},\
  }\href@noop {} {\bibfield  {journal} {\bibinfo  {journal} {Nature Materials}\
  }\textbf {\bibinfo {volume} {19}},\ \bibinfo {pages} {1164–1168} (\bibinfo
  {year} {2020})}\BibitemShut {NoStop}%
\bibitem [{\citenamefont {Pelc}\ \emph {et~al.}(2012)\citenamefont {Pelc},
  \citenamefont {Yu}, \citenamefont {De~Greve}, \citenamefont {McMahon},
  \citenamefont {Natarajan}, \citenamefont {Esfandyarpour}, \citenamefont
  {Maier}, \citenamefont {Schneider}, \citenamefont {Kamp}, \citenamefont
  {H\"ofling}, \citenamefont {Hadfield}, \citenamefont {Forchel}, \citenamefont
  {Yamamoto},\ and\ \citenamefont {Fejer}}]{Pelc2012}%
  \BibitemOpen
  \bibfield  {author} {\bibinfo {author} {\bibfnamefont {J.}~\bibnamefont
  {Pelc}}, \bibinfo {author} {\bibfnamefont {L.}~\bibnamefont {Yu}}, \bibinfo
  {author} {\bibfnamefont {K.}~\bibnamefont {De~Greve}}, \bibinfo {author}
  {\bibfnamefont {P.}~\bibnamefont {McMahon}}, \bibinfo {author} {\bibfnamefont
  {C.}~\bibnamefont {Natarajan}}, \bibinfo {author} {\bibfnamefont
  {V.}~\bibnamefont {Esfandyarpour}}, \bibinfo {author} {\bibfnamefont
  {S.}~\bibnamefont {Maier}}, \bibinfo {author} {\bibfnamefont
  {C.}~\bibnamefont {Schneider}}, \bibinfo {author} {\bibfnamefont
  {M.}~\bibnamefont {Kamp}}, \bibinfo {author} {\bibfnamefont {S.}~\bibnamefont
  {H\"ofling}}, \bibinfo {author} {\bibfnamefont {R.}~\bibnamefont {Hadfield}},
  \bibinfo {author} {\bibfnamefont {A.}~\bibnamefont {Forchel}}, \bibinfo
  {author} {\bibfnamefont {Y.}~\bibnamefont {Yamamoto}},\ and\ \bibinfo
  {author} {\bibfnamefont {M.}~\bibnamefont {Fejer}},\ }\bibfield  {title}
  {\bibinfo {title} {Downconversion quantum interface for a single quantum dot
  spin and 1550-nm single-photon channel},\ }\href@noop {} {\bibfield
  {journal} {\bibinfo  {journal} {Optics Express}\ }\textbf {\bibinfo {volume}
  {20}},\ \bibinfo {pages} {27510} (\bibinfo {year} {2012})}\BibitemShut
  {NoStop}%
\bibitem [{\citenamefont {Kambs}\ \emph {et~al.}(2016)\citenamefont {Kambs},
  \citenamefont {Kettler}, \citenamefont {Bock}, \citenamefont {Becker},
  \citenamefont {Arend}, \citenamefont {Lenhard}, \citenamefont {Portalupi},
  \citenamefont {Jetter}, \citenamefont {Michler},\ and\ \citenamefont
  {Becher}}]{Kambs2016}%
  \BibitemOpen
  \bibfield  {author} {\bibinfo {author} {\bibfnamefont {B.}~\bibnamefont
  {Kambs}}, \bibinfo {author} {\bibfnamefont {J.}~\bibnamefont {Kettler}},
  \bibinfo {author} {\bibfnamefont {M.}~\bibnamefont {Bock}}, \bibinfo {author}
  {\bibfnamefont {J.}~\bibnamefont {Becker}}, \bibinfo {author} {\bibfnamefont
  {C.}~\bibnamefont {Arend}}, \bibinfo {author} {\bibfnamefont
  {A.}~\bibnamefont {Lenhard}}, \bibinfo {author} {\bibfnamefont
  {S.}~\bibnamefont {Portalupi}}, \bibinfo {author} {\bibfnamefont
  {M.}~\bibnamefont {Jetter}}, \bibinfo {author} {\bibfnamefont
  {P.}~\bibnamefont {Michler}},\ and\ \bibinfo {author} {\bibfnamefont
  {C.}~\bibnamefont {Becher}},\ }\bibfield  {title} {\bibinfo {title}
  {Low-noise quantum frequency down-conversion of indistinguishable photons},\
  }\href@noop {} {\bibfield  {journal} {\bibinfo  {journal} {Optics Express}\
  }\textbf {\bibinfo {volume} {24}},\ \bibinfo {pages} {22250} (\bibinfo {year}
  {2016})}\BibitemShut {NoStop}%
\bibitem [{\citenamefont {Da~Lio}\ \emph {et~al.}(2022)\citenamefont {Da~Lio},
  \citenamefont {Faurby}, \citenamefont {Zhou}, \citenamefont {Chan},
  \citenamefont {Uppu}, \citenamefont {Thyrrestrup}, \citenamefont {Scholz},
  \citenamefont {Wieck}, \citenamefont {Ludwig}, \citenamefont {Lodahl},\ and\
  \citenamefont {Midolo}}]{Lio2022}%
  \BibitemOpen
  \bibfield  {author} {\bibinfo {author} {\bibfnamefont {B.}~\bibnamefont
  {Da~Lio}}, \bibinfo {author} {\bibfnamefont {C.}~\bibnamefont {Faurby}},
  \bibinfo {author} {\bibfnamefont {X.}~\bibnamefont {Zhou}}, \bibinfo {author}
  {\bibfnamefont {M.}~\bibnamefont {Chan}}, \bibinfo {author} {\bibfnamefont
  {R.}~\bibnamefont {Uppu}}, \bibinfo {author} {\bibfnamefont {H.}~\bibnamefont
  {Thyrrestrup}}, \bibinfo {author} {\bibfnamefont {S.}~\bibnamefont {Scholz}},
  \bibinfo {author} {\bibfnamefont {A.}~\bibnamefont {Wieck}}, \bibinfo
  {author} {\bibfnamefont {A.}~\bibnamefont {Ludwig}}, \bibinfo {author}
  {\bibfnamefont {P.}~\bibnamefont {Lodahl}},\ and\ \bibinfo {author}
  {\bibfnamefont {L.}~\bibnamefont {Midolo}},\ }\bibfield  {title} {\bibinfo
  {title} {A pure and indistinguishable single-photon source at
  telecommunication wavelength},\ }\href@noop {} {\bibfield  {journal}
  {\bibinfo  {journal} {Advanced Quantum Technologies}\ }\textbf {\bibinfo
  {volume} {5}},\ \bibinfo {pages} {2200006} (\bibinfo {year}
  {2022})}\BibitemShut {NoStop}%
\bibitem [{\citenamefont {Morrison}\ \emph {et~al.}(2021)\citenamefont
  {Morrison}, \citenamefont {Rambach}, \citenamefont {Koong}, \citenamefont
  {Graffitti}, \citenamefont {Thorburn}, \citenamefont {Kar}, \citenamefont
  {Ma}, \citenamefont {Park}, \citenamefont {Song}, \citenamefont {Stoltz},
  \citenamefont {Bouwmeester}, \citenamefont {Fedrizzi},\ and\ \citenamefont
  {Gerardot}}]{Morrison2021}%
  \BibitemOpen
  \bibfield  {author} {\bibinfo {author} {\bibfnamefont {C.~L.}\ \bibnamefont
  {Morrison}}, \bibinfo {author} {\bibfnamefont {M.}~\bibnamefont {Rambach}},
  \bibinfo {author} {\bibfnamefont {Z.~X.}\ \bibnamefont {Koong}}, \bibinfo
  {author} {\bibfnamefont {F.}~\bibnamefont {Graffitti}}, \bibinfo {author}
  {\bibfnamefont {F.}~\bibnamefont {Thorburn}}, \bibinfo {author}
  {\bibfnamefont {A.~K.}\ \bibnamefont {Kar}}, \bibinfo {author} {\bibfnamefont
  {Y.}~\bibnamefont {Ma}}, \bibinfo {author} {\bibfnamefont {S.-I.}\
  \bibnamefont {Park}}, \bibinfo {author} {\bibfnamefont {J.~D.}\ \bibnamefont
  {Song}}, \bibinfo {author} {\bibfnamefont {N.~G.}\ \bibnamefont {Stoltz}},
  \bibinfo {author} {\bibfnamefont {D.}~\bibnamefont {Bouwmeester}}, \bibinfo
  {author} {\bibfnamefont {A.}~\bibnamefont {Fedrizzi}},\ and\ \bibinfo
  {author} {\bibfnamefont {B.~D.}\ \bibnamefont {Gerardot}},\ }\bibfield
  {title} {\bibinfo {title} {A bright source of telecom single photons based on
  quantum frequency conversion},\ }\href@noop {} {\bibfield  {journal}
  {\bibinfo  {journal} {Applied Physics Letters}\ }\textbf {\bibinfo {volume}
  {118}},\ \bibinfo {pages} {174003} (\bibinfo {year} {2021})}\BibitemShut
  {NoStop}%
\bibitem [{\citenamefont {Singh}\ \emph {et~al.}(2019)\citenamefont {Singh},
  \citenamefont {Li}, \citenamefont {Liu}, \citenamefont {Yu}, \citenamefont
  {Lu}, \citenamefont {Schneider}, \citenamefont {H\"ofling}, \citenamefont
  {Lawall}, \citenamefont {Verma}, \citenamefont {Mirin}, \citenamefont {Nam},
  \citenamefont {Liu},\ and\ \citenamefont {Srinivasan}}]{Singh2019}%
  \BibitemOpen
  \bibfield  {author} {\bibinfo {author} {\bibfnamefont {A.}~\bibnamefont
  {Singh}}, \bibinfo {author} {\bibfnamefont {Q.}~\bibnamefont {Li}}, \bibinfo
  {author} {\bibfnamefont {S.}~\bibnamefont {Liu}}, \bibinfo {author}
  {\bibfnamefont {Y.}~\bibnamefont {Yu}}, \bibinfo {author} {\bibfnamefont
  {X.}~\bibnamefont {Lu}}, \bibinfo {author} {\bibfnamefont {C.}~\bibnamefont
  {Schneider}}, \bibinfo {author} {\bibfnamefont {S.}~\bibnamefont
  {H\"ofling}}, \bibinfo {author} {\bibfnamefont {J.}~\bibnamefont {Lawall}},
  \bibinfo {author} {\bibfnamefont {V.}~\bibnamefont {Verma}}, \bibinfo
  {author} {\bibfnamefont {R.}~\bibnamefont {Mirin}}, \bibinfo {author}
  {\bibfnamefont {S.}~\bibnamefont {Nam}}, \bibinfo {author} {\bibfnamefont
  {J.}~\bibnamefont {Liu}},\ and\ \bibinfo {author} {\bibfnamefont
  {K.}~\bibnamefont {Srinivasan}},\ }\bibfield  {title} {\bibinfo {title}
  {Quantum frequency conversion of a quantum dot single-photon source on a
  nanophotonic chip},\ }\href@noop {} {\bibfield  {journal} {\bibinfo
  {journal} {Optica}\ }\textbf {\bibinfo {volume} {6}},\ \bibinfo {pages} {563}
  (\bibinfo {year} {2019})}\BibitemShut {NoStop}%
\bibitem [{\citenamefont {Yatsiv}\ \emph {et~al.}(1968)\citenamefont {Yatsiv},
  \citenamefont {Rokni},\ and\ \citenamefont {Barak}}]{Yatsiv1968}%
  \BibitemOpen
  \bibfield  {author} {\bibinfo {author} {\bibfnamefont {S.}~\bibnamefont
  {Yatsiv}}, \bibinfo {author} {\bibfnamefont {M.}~\bibnamefont {Rokni}},\ and\
  \bibinfo {author} {\bibfnamefont {S.}~\bibnamefont {Barak}},\ }\bibfield
  {title} {\bibinfo {title} {Enhanced two-proton emission},\ }\href
  {https://doi.org/10.1103/PhysRevLett.20.1282} {\bibfield  {journal} {\bibinfo
   {journal} {Phys. Rev. Lett.}\ }\textbf {\bibinfo {volume} {20}},\ \bibinfo
  {pages} {1282} (\bibinfo {year} {1968})}\BibitemShut {NoStop}%
\bibitem [{\citenamefont {Hennrich}\ \emph {et~al.}(2000)\citenamefont
  {Hennrich}, \citenamefont {Legero}, \citenamefont {Kuhn},\ and\ \citenamefont
  {Rempe}}]{Hennrich2000}%
  \BibitemOpen
  \bibfield  {author} {\bibinfo {author} {\bibfnamefont {M.}~\bibnamefont
  {Hennrich}}, \bibinfo {author} {\bibfnamefont {T.}~\bibnamefont {Legero}},
  \bibinfo {author} {\bibfnamefont {A.}~\bibnamefont {Kuhn}},\ and\ \bibinfo
  {author} {\bibfnamefont {G.}~\bibnamefont {Rempe}},\ }\bibfield  {title}
  {\bibinfo {title} {Vacuum-stimulated raman scattering based on adiabatic
  passage in a high-finesse optical cavity},\ }\href
  {https://doi.org/10.1103/PhysRevLett.85.4872} {\bibfield  {journal} {\bibinfo
   {journal} {Physical Review Letters}\ }\textbf {\bibinfo {volume} {85}},\
  \bibinfo {pages} {4872} (\bibinfo {year} {2000})}\BibitemShut {NoStop}%
\bibitem [{\citenamefont {Breddermann}\ \emph {et~al.}(2016)\citenamefont
  {Breddermann}, \citenamefont {Heinze}, \citenamefont {Binder}, \citenamefont
  {Zrenner},\ and\ \citenamefont {Schumacher}}]{Schumacher2016}%
  \BibitemOpen
  \bibfield  {author} {\bibinfo {author} {\bibfnamefont {D.}~\bibnamefont
  {Breddermann}}, \bibinfo {author} {\bibfnamefont {D.}~\bibnamefont {Heinze}},
  \bibinfo {author} {\bibfnamefont {R.}~\bibnamefont {Binder}}, \bibinfo
  {author} {\bibfnamefont {A.}~\bibnamefont {Zrenner}},\ and\ \bibinfo {author}
  {\bibfnamefont {S.}~\bibnamefont {Schumacher}},\ }\bibfield  {title}
  {\bibinfo {title} {All-optical tailoring of single-photon spectra in a
  quantum-dot microcavity system},\ }\href
  {https://doi.org/10.1103/PhysRevB.94.165310} {\bibfield  {journal} {\bibinfo
  {journal} {Physical Review B}\ }\textbf {\bibinfo {volume} {94}},\ \bibinfo
  {pages} {165310} (\bibinfo {year} {2016})}\BibitemShut {NoStop}%
\bibitem [{\citenamefont {Praschan}\ \emph {et~al.}(2022)\citenamefont
  {Praschan}, \citenamefont {Heinze}, \citenamefont {Breddermann},
  \citenamefont {Zrenner}, \citenamefont {Walther},\ and\ \citenamefont
  {Schumacher}}]{Schumacher2022}%
  \BibitemOpen
  \bibfield  {author} {\bibinfo {author} {\bibfnamefont {T.}~\bibnamefont
  {Praschan}}, \bibinfo {author} {\bibfnamefont {D.}~\bibnamefont {Heinze}},
  \bibinfo {author} {\bibfnamefont {D.}~\bibnamefont {Breddermann}}, \bibinfo
  {author} {\bibfnamefont {A.}~\bibnamefont {Zrenner}}, \bibinfo {author}
  {\bibfnamefont {A.}~\bibnamefont {Walther}},\ and\ \bibinfo {author}
  {\bibfnamefont {S.}~\bibnamefont {Schumacher}},\ }\bibfield  {title}
  {\bibinfo {title} {Pulse shaping for on-demand emission of single raman
  photons from a quantum-dot biexciton},\ }\href
  {https://doi.org/10.1103/PhysRevB.105.045302} {\bibfield  {journal} {\bibinfo
   {journal} {Physical Review B}\ }\textbf {\bibinfo {volume} {105}},\ \bibinfo
  {pages} {045302} (\bibinfo {year} {2022})}\BibitemShut {NoStop}%
\bibitem [{\citenamefont {Jonas}\ \emph {et~al.}(2022)\citenamefont {Jonas},
  \citenamefont {Heinze}, \citenamefont {Sch{\"o}ll}, \citenamefont {Kallert},
  \citenamefont {Langer}, \citenamefont {Krehs}, \citenamefont {Widhalm},
  \citenamefont {J{\"o}ns}, \citenamefont {Reuter}, \citenamefont
  {Schumacher},\ and\ \citenamefont {Zrenner}}]{Jonas2022}%
  \BibitemOpen
  \bibfield  {author} {\bibinfo {author} {\bibfnamefont {B.}~\bibnamefont
  {Jonas}}, \bibinfo {author} {\bibfnamefont {D.}~\bibnamefont {Heinze}},
  \bibinfo {author} {\bibfnamefont {E.}~\bibnamefont {Sch{\"o}ll}}, \bibinfo
  {author} {\bibfnamefont {P.}~\bibnamefont {Kallert}}, \bibinfo {author}
  {\bibfnamefont {T.}~\bibnamefont {Langer}}, \bibinfo {author} {\bibfnamefont
  {S.}~\bibnamefont {Krehs}}, \bibinfo {author} {\bibfnamefont
  {A.}~\bibnamefont {Widhalm}}, \bibinfo {author} {\bibfnamefont {K.~D.}\
  \bibnamefont {J{\"o}ns}}, \bibinfo {author} {\bibfnamefont {D.}~\bibnamefont
  {Reuter}}, \bibinfo {author} {\bibfnamefont {S.}~\bibnamefont {Schumacher}},\
  and\ \bibinfo {author} {\bibfnamefont {A.}~\bibnamefont {Zrenner}},\
  }\bibfield  {title} {\bibinfo {title} {Nonlinear down-conversion in a single
  quantum dot},\ }\href {https://doi.org/10.1038/s41467-022-28993-3} {\bibfield
   {journal} {\bibinfo  {journal} {Nature Communications}\ }\textbf {\bibinfo
  {volume} {13}},\ \bibinfo {pages} {1387} (\bibinfo {year}
  {2022})}\BibitemShut {NoStop}%
\bibitem [{\citenamefont {Wei}\ \emph {et~al.}(2022)\citenamefont {Wei},
  \citenamefont {Liu}, \citenamefont {Li}, \citenamefont {Yu}, \citenamefont
  {Su}, \citenamefont {Li}, \citenamefont {Shang}, \citenamefont {Liu},
  \citenamefont {Hao}, \citenamefont {Ni}, \citenamefont {Yu}, \citenamefont
  {Niu}, \citenamefont {Iles-Smith}, \citenamefont {Liu},\ and\ \citenamefont
  {Wang}}]{Wei2022}%
  \BibitemOpen
  \bibfield  {author} {\bibinfo {author} {\bibfnamefont {Y.}~\bibnamefont
  {Wei}}, \bibinfo {author} {\bibfnamefont {S.}~\bibnamefont {Liu}}, \bibinfo
  {author} {\bibfnamefont {X.}~\bibnamefont {Li}}, \bibinfo {author}
  {\bibfnamefont {Y.}~\bibnamefont {Yu}}, \bibinfo {author} {\bibfnamefont
  {X.}~\bibnamefont {Su}}, \bibinfo {author} {\bibfnamefont {S.}~\bibnamefont
  {Li}}, \bibinfo {author} {\bibfnamefont {X.}~\bibnamefont {Shang}}, \bibinfo
  {author} {\bibfnamefont {H.}~\bibnamefont {Liu}}, \bibinfo {author}
  {\bibfnamefont {H.}~\bibnamefont {Hao}}, \bibinfo {author} {\bibfnamefont
  {H.}~\bibnamefont {Ni}}, \bibinfo {author} {\bibfnamefont {S.}~\bibnamefont
  {Yu}}, \bibinfo {author} {\bibfnamefont {Z.}~\bibnamefont {Niu}}, \bibinfo
  {author} {\bibfnamefont {J.}~\bibnamefont {Iles-Smith}}, \bibinfo {author}
  {\bibfnamefont {J.}~\bibnamefont {Liu}},\ and\ \bibinfo {author}
  {\bibfnamefont {X.}~\bibnamefont {Wang}},\ }\bibfield  {title} {\bibinfo
  {title} {Tailoring solid-state single-photon sources with stimulated
  emissions},\ }\href@noop {} {\bibfield  {journal} {\bibinfo  {journal}
  {Nature Nanotechnology}\ }\textbf {\bibinfo {volume} {17}},\ \bibinfo {pages}
  {470–476} (\bibinfo {year} {2022})}\BibitemShut {NoStop}%
\bibitem [{\citenamefont {Schwartz}\ \emph {et~al.}(2015)\citenamefont
  {Schwartz}, \citenamefont {Schmidgall}, \citenamefont {Gantz}, \citenamefont
  {Cogan}, \citenamefont {Bordo}, \citenamefont {Don}, \citenamefont
  {Zielinski},\ and\ \citenamefont {Gershoni}}]{Schawrtz2015}%
  \BibitemOpen
  \bibfield  {author} {\bibinfo {author} {\bibfnamefont {I.}~\bibnamefont
  {Schwartz}}, \bibinfo {author} {\bibfnamefont {E.~R.}\ \bibnamefont
  {Schmidgall}}, \bibinfo {author} {\bibfnamefont {L.}~\bibnamefont {Gantz}},
  \bibinfo {author} {\bibfnamefont {D.}~\bibnamefont {Cogan}}, \bibinfo
  {author} {\bibfnamefont {E.}~\bibnamefont {Bordo}}, \bibinfo {author}
  {\bibfnamefont {Y.}~\bibnamefont {Don}}, \bibinfo {author} {\bibfnamefont
  {M.}~\bibnamefont {Zielinski}},\ and\ \bibinfo {author} {\bibfnamefont
  {D.}~\bibnamefont {Gershoni}},\ }\bibfield  {title} {\bibinfo {title}
  {Deterministic writing and control of the dark exciton spin using single
  short optical pulses},\ }\href@noop {} {\bibfield  {journal} {\bibinfo
  {journal} {Physical Review X}\ }\textbf {\bibinfo {volume} {5}},\ \bibinfo
  {pages} {011009} (\bibinfo {year} {2015})}\BibitemShut {NoStop}%
\bibitem [{\citenamefont {Musial}\ \emph {et~al.}(2021)\citenamefont {Musial},
  \citenamefont {Mikulicz}, \citenamefont {Mrowi\'nski}, \citenamefont
  {Zieli\'nska}, \citenamefont {Sitarek}, \citenamefont {Wyborski},
  \citenamefont {Kuniej}, \citenamefont {Reithmaier}, \citenamefont {Sek},\
  and\ \citenamefont {Benyoucef}}]{Musial2021}%
  \BibitemOpen
  \bibfield  {author} {\bibinfo {author} {\bibfnamefont {A.}~\bibnamefont
  {Musial}}, \bibinfo {author} {\bibfnamefont {M.}~\bibnamefont {Mikulicz}},
  \bibinfo {author} {\bibfnamefont {P.}~\bibnamefont {Mrowi\'nski}}, \bibinfo
  {author} {\bibfnamefont {A.}~\bibnamefont {Zieli\'nska}}, \bibinfo {author}
  {\bibfnamefont {P.}~\bibnamefont {Sitarek}}, \bibinfo {author} {\bibfnamefont
  {P.}~\bibnamefont {Wyborski}}, \bibinfo {author} {\bibfnamefont
  {M.}~\bibnamefont {Kuniej}}, \bibinfo {author} {\bibfnamefont {J.~P.}\
  \bibnamefont {Reithmaier}}, \bibinfo {author} {\bibfnamefont
  {G.}~\bibnamefont {Sek}},\ and\ \bibinfo {author} {\bibfnamefont
  {M.}~\bibnamefont {Benyoucef}},\ }\bibfield  {title} {\bibinfo {title}
  {{In{P}-based single-photon sources operating at telecom {C}-band with
  increased extraction efficiency}},\ }\href@noop {} {\bibfield  {journal}
  {\bibinfo  {journal} {Applied Physics Letters}\ }\textbf {\bibinfo {volume}
  {118}},\ \bibinfo {pages} {221101} (\bibinfo {year} {2021})}\BibitemShut
  {NoStop}%
\bibitem [{\citenamefont {Wells}\ \emph {et~al.}(2023)\citenamefont {Wells},
  \citenamefont {M\"uller}, \citenamefont {Stevenson}, \citenamefont
  {Skiba-Szymanska}, \citenamefont {Ritchie},\ and\ \citenamefont
  {Shields}}]{Wells2023}%
  \BibitemOpen
  \bibfield  {author} {\bibinfo {author} {\bibfnamefont {L.}~\bibnamefont
  {Wells}}, \bibinfo {author} {\bibfnamefont {T.}~\bibnamefont {M\"uller}},
  \bibinfo {author} {\bibfnamefont {R.}~\bibnamefont {Stevenson}}, \bibinfo
  {author} {\bibfnamefont {J.}~\bibnamefont {Skiba-Szymanska}}, \bibinfo
  {author} {\bibfnamefont {D.}~\bibnamefont {Ritchie}},\ and\ \bibinfo {author}
  {\bibfnamefont {A.}~\bibnamefont {Shields}},\ }\bibfield  {title} {\bibinfo
  {title} {Coherent light scattering from a telecom c-band quantum dot},\
  }\href@noop {} {\bibfield  {journal} {\bibinfo  {journal} {Nature
  Communications}\ }\textbf {\bibinfo {volume} {14}},\ \bibinfo {pages} {8371}
  (\bibinfo {year} {2023})}\BibitemShut {NoStop}%
\bibitem [{\citenamefont {Schneider}\ \emph {et~al.}(2016)\citenamefont
  {Schneider}, \citenamefont {Gold}, \citenamefont {Reitzenstein},
  \citenamefont {H\"ofling},\ and\ \citenamefont {Kamp}}]{Schneider2016}%
  \BibitemOpen
  \bibfield  {author} {\bibinfo {author} {\bibfnamefont {C.}~\bibnamefont
  {Schneider}}, \bibinfo {author} {\bibfnamefont {P.}~\bibnamefont {Gold}},
  \bibinfo {author} {\bibfnamefont {S.}~\bibnamefont {Reitzenstein}}, \bibinfo
  {author} {\bibfnamefont {S.}~\bibnamefont {H\"ofling}},\ and\ \bibinfo
  {author} {\bibfnamefont {M.}~\bibnamefont {Kamp}},\ }\bibfield  {title}
  {\bibinfo {title} {Quantum dot micropillar cavities with quality factors
  exceeding 250000},\ }\href@noop {} {\bibfield  {journal} {\bibinfo  {journal}
  {Applied Physics B}\ }\textbf {\bibinfo {volume} {122}},\ \bibinfo {pages}
  {19} (\bibinfo {year} {2016})}\BibitemShut {NoStop}%
\bibitem [{\citenamefont {Kitamura}\ \emph {et~al.}(2015)\citenamefont
  {Kitamura}, \citenamefont {Senshu}, \citenamefont {Katsuyama}, \citenamefont
  {Hino}, \citenamefont {Ozaki}, \citenamefont {Ohkouchi}, \citenamefont
  {Sugimoto},\ and\ \citenamefont {Hogg}}]{Kitamura2015}%
  \BibitemOpen
  \bibfield  {author} {\bibinfo {author} {\bibfnamefont {S.}~\bibnamefont
  {Kitamura}}, \bibinfo {author} {\bibfnamefont {M.}~\bibnamefont {Senshu}},
  \bibinfo {author} {\bibfnamefont {T.}~\bibnamefont {Katsuyama}}, \bibinfo
  {author} {\bibfnamefont {Y.}~\bibnamefont {Hino}}, \bibinfo {author}
  {\bibfnamefont {N.}~\bibnamefont {Ozaki}}, \bibinfo {author} {\bibfnamefont
  {S.}~\bibnamefont {Ohkouchi}}, \bibinfo {author} {\bibfnamefont
  {Y.}~\bibnamefont {Sugimoto}},\ and\ \bibinfo {author} {\bibfnamefont
  {R.}~\bibnamefont {Hogg}},\ }\bibfield  {title} {\bibinfo {title} {Optical
  characterization of in-flushed {I}n{A}s/{G}a{A}s quantum dots emitting a
  broadband spectrum with multiple peaks at $\sim$1 $\mu$m},\ }\href@noop {}
  {\bibfield  {journal} {\bibinfo  {journal} {Nano Express}\ }\textbf {\bibinfo
  {volume} {10}},\ \bibinfo {pages} {231} (\bibinfo {year} {2015})}\BibitemShut
  {NoStop}%
\bibitem [{\citenamefont {Hu}\ \emph {et~al.}(2020)\citenamefont {Hu},
  \citenamefont {Zhang}, \citenamefont {Guzun}, \citenamefont {Ware},
  \citenamefont {Mazur}, \citenamefont {Lienau},\ and\ \citenamefont
  {Salamo}}]{Hu2020}%
  \BibitemOpen
  \bibfield  {author} {\bibinfo {author} {\bibfnamefont {X.}~\bibnamefont
  {Hu}}, \bibinfo {author} {\bibfnamefont {Y.}~\bibnamefont {Zhang}}, \bibinfo
  {author} {\bibfnamefont {D.}~\bibnamefont {Guzun}}, \bibinfo {author}
  {\bibfnamefont {M.}~\bibnamefont {Ware}}, \bibinfo {author} {\bibfnamefont
  {Y.}~\bibnamefont {Mazur}}, \bibinfo {author} {\bibfnamefont
  {C.}~\bibnamefont {Lienau}},\ and\ \bibinfo {author} {\bibfnamefont
  {G.}~\bibnamefont {Salamo}},\ }\bibfield  {title} {\bibinfo {title}
  {Photoluminescence of {I}n{A}s/{G}a{A}s quantum dots under direct two-photon
  excitation},\ }\href@noop {} {\bibfield  {journal} {\bibinfo  {journal}
  {Scientific Reports}\ }\textbf {\bibinfo {volume} {10}},\ \bibinfo {pages}
  {10930} (\bibinfo {year} {2020})}\BibitemShut {NoStop}%
\bibitem [{\citenamefont {Durnev}\ \emph {et~al.}(2014)\citenamefont {Durnev},
  \citenamefont {Glazov},\ and\ \citenamefont {Ivchenko}}]{Ivchenko_2014}%
  \BibitemOpen
  \bibfield  {author} {\bibinfo {author} {\bibfnamefont {M.~V.}\ \bibnamefont
  {Durnev}}, \bibinfo {author} {\bibfnamefont {M.~M.}\ \bibnamefont {Glazov}},\
  and\ \bibinfo {author} {\bibfnamefont {E.~L.}\ \bibnamefont {Ivchenko}},\
  }\bibfield  {title} {\bibinfo {title} {Spin-orbit splitting of valence
  subbands in semiconductor nanostructures},\ }\href
  {https://doi.org/10.1103/PhysRevB.89.075430} {\bibfield  {journal} {\bibinfo
  {journal} {Phys. Rev. B}\ }\textbf {\bibinfo {volume} {89}},\ \bibinfo
  {pages} {075430} (\bibinfo {year} {2014})}\BibitemShut {NoStop}%
\bibitem [{\citenamefont {{In Lee}}\ \emph {et~al.}(1998)\citenamefont {{In
  Lee}}, \citenamefont {{Gyoo Lee}}, \citenamefont {Shin}, \citenamefont {Yu},
  \citenamefont {Viswanath}, \citenamefont {Kim},\ and\ \citenamefont
  {Ihm}}]{Lee1998}%
  \BibitemOpen
  \bibfield  {author} {\bibinfo {author} {\bibfnamefont {J.}~\bibnamefont {{In
  Lee}}}, \bibinfo {author} {\bibfnamefont {H.}~\bibnamefont {{Gyoo Lee}}},
  \bibinfo {author} {\bibfnamefont {E.-J.}\ \bibnamefont {Shin}}, \bibinfo
  {author} {\bibfnamefont {S.}~\bibnamefont {Yu}}, \bibinfo {author}
  {\bibfnamefont {K.}~\bibnamefont {Viswanath}}, \bibinfo {author}
  {\bibfnamefont {D.}~\bibnamefont {Kim}},\ and\ \bibinfo {author}
  {\bibfnamefont {G.}~\bibnamefont {Ihm}},\ }\bibfield  {title} {\bibinfo
  {title} {Time-resolved spectroscopy of {I}n{A}s quantum dots using one-side
  modulation-doping technique: renormalization and screening},\ }\href
  {https://doi.org/https://doi.org/10.1016/S0921-5107(97)00244-4} {\bibfield
  {journal} {\bibinfo  {journal} {Materials Science and Engineering: B}\
  }\textbf {\bibinfo {volume} {51}},\ \bibinfo {pages} {122} (\bibinfo {year}
  {1998})}\BibitemShut {NoStop}%
\bibitem [{\citenamefont {Kono}\ \emph {et~al.}(2005)\citenamefont {Kono},
  \citenamefont {Kirihara}, \citenamefont {Tomita}, \citenamefont {Nakamura},
  \citenamefont {Fujikata}, \citenamefont {Ohashi}, \citenamefont {Saito},\
  and\ \citenamefont {Nishi}}]{Kono2005}%
  \BibitemOpen
  \bibfield  {author} {\bibinfo {author} {\bibfnamefont {S.}~\bibnamefont
  {Kono}}, \bibinfo {author} {\bibfnamefont {A.}~\bibnamefont {Kirihara}},
  \bibinfo {author} {\bibfnamefont {A.}~\bibnamefont {Tomita}}, \bibinfo
  {author} {\bibfnamefont {K.}~\bibnamefont {Nakamura}}, \bibinfo {author}
  {\bibfnamefont {J.}~\bibnamefont {Fujikata}}, \bibinfo {author}
  {\bibfnamefont {K.}~\bibnamefont {Ohashi}}, \bibinfo {author} {\bibfnamefont
  {H.}~\bibnamefont {Saito}},\ and\ \bibinfo {author} {\bibfnamefont
  {K.}~\bibnamefont {Nishi}},\ }\bibfield  {title} {\bibinfo {title} {Excitonic
  molecule in a quantum dot: Photoluminescence lifetime of a single
  $\mathrm{In}\mathrm{As}/ \mathrm{Ga}\mathrm{As}$ quantum dot},\ }\href@noop
  {} {\bibfield  {journal} {\bibinfo  {journal} {Physical Review B}\ }\textbf
  {\bibinfo {volume} {72}},\ \bibinfo {pages} {155307} (\bibinfo {year}
  {2005})}\BibitemShut {NoStop}%
\bibitem [{\citenamefont {Holewa}\ \emph {et~al.}(2020)\citenamefont {Holewa},
  \citenamefont {Gawe\l{}czyk}, \citenamefont {Ciostek}, \citenamefont
  {Wyborski}, \citenamefont {Kadkhodazadeh}, \citenamefont {Semenova},\ and\
  \citenamefont {Syperek}}]{Holewa2020}%
  \BibitemOpen
  \bibfield  {author} {\bibinfo {author} {\bibfnamefont {P.}~\bibnamefont
  {Holewa}}, \bibinfo {author} {\bibfnamefont {M.}~\bibnamefont
  {Gawe\l{}czyk}}, \bibinfo {author} {\bibfnamefont {C.}~\bibnamefont
  {Ciostek}}, \bibinfo {author} {\bibfnamefont {P.}~\bibnamefont {Wyborski}},
  \bibinfo {author} {\bibfnamefont {S.}~\bibnamefont {Kadkhodazadeh}}, \bibinfo
  {author} {\bibfnamefont {E.}~\bibnamefont {Semenova}},\ and\ \bibinfo
  {author} {\bibfnamefont {M.}~\bibnamefont {Syperek}},\ }\bibfield  {title}
  {\bibinfo {title} {Optical and electronic properties of low-density
  {I}n{A}s/{I}n{P} quantum-dot-like structures designed for single-photon
  emitters at telecom wavelengths},\ }\href@noop {} {\bibfield  {journal}
  {\bibinfo  {journal} {Physical Review X}\ }\textbf {\bibinfo {volume}
  {101}},\ \bibinfo {pages} {195304} (\bibinfo {year} {2020})}\BibitemShut
  {NoStop}%
\bibitem [{\citenamefont {Wyborski}\ \emph {et~al.}(2023)\citenamefont
  {Wyborski}, \citenamefont {Gawe\l{}czyk}, \citenamefont {Podemski},
  \citenamefont {Wro\'nski}, \citenamefont {Pawlyta}, \citenamefont {Gorantla},
  \citenamefont {Jabeen}, \citenamefont {H\"ofling},\ and\ \citenamefont
  {Sek}}]{Wyborski2023}%
  \BibitemOpen
  \bibfield  {author} {\bibinfo {author} {\bibfnamefont {P.}~\bibnamefont
  {Wyborski}}, \bibinfo {author} {\bibfnamefont {M.}~\bibnamefont
  {Gawe\l{}czyk}}, \bibinfo {author} {\bibfnamefont {P.}~\bibnamefont
  {Podemski}}, \bibinfo {author} {\bibfnamefont {P.}~\bibnamefont {Wro\'nski}},
  \bibinfo {author} {\bibfnamefont {M.}~\bibnamefont {Pawlyta}}, \bibinfo
  {author} {\bibfnamefont {S.}~\bibnamefont {Gorantla}}, \bibinfo {author}
  {\bibfnamefont {F.}~\bibnamefont {Jabeen}}, \bibinfo {author} {\bibfnamefont
  {S.}~\bibnamefont {H\"ofling}},\ and\ \bibinfo {author} {\bibfnamefont
  {G.}~\bibnamefont {Sek}},\ }\bibfield  {title} {\bibinfo {title} {Impact of
  {MBE}-grown $(\mathrm{In},\mathrm{Ga})\mathrm{As}/\mathrm{Ga}\mathrm{As}$
  metamorphic buffers on excitonic and optical properties of single quantum
  dots with single-photon emission tuned to the telecom range},\ }\href@noop {}
  {\bibfield  {journal} {\bibinfo  {journal} {Physical Review Applied}\
  }\textbf {\bibinfo {volume} {20}},\ \bibinfo {pages} {044009} (\bibinfo
  {year} {2023})}\BibitemShut {NoStop}%
\bibitem [{\citenamefont {Holewa}\ \emph {et~al.}(2024)\citenamefont {Holewa},
  \citenamefont {Vajner}, \citenamefont {Zieba-Ost\'oj}, \citenamefont
  {Wasiluk}, \citenamefont {Ga\'al}, \citenamefont {Sakanas}, \citenamefont
  {Burakowski}, \citenamefont {Mrowi\'nski}, \citenamefont {Krajnik},
  \citenamefont {Xiong}, \citenamefont {Yvind}, \citenamefont {Gregersen},
  \citenamefont {Musial}, \citenamefont {Huck}, \citenamefont {Heindel},
  \citenamefont {Syperek},\ and\ \citenamefont {Semenova}}]{Holewa2024}%
  \BibitemOpen
  \bibfield  {author} {\bibinfo {author} {\bibfnamefont {P.}~\bibnamefont
  {Holewa}}, \bibinfo {author} {\bibfnamefont {D.}~\bibnamefont {Vajner}},
  \bibinfo {author} {\bibfnamefont {E.}~\bibnamefont {Zieba-Ost\'oj}}, \bibinfo
  {author} {\bibfnamefont {M.}~\bibnamefont {Wasiluk}}, \bibinfo {author}
  {\bibfnamefont {B.}~\bibnamefont {Ga\'al}}, \bibinfo {author} {\bibfnamefont
  {A.}~\bibnamefont {Sakanas}}, \bibinfo {author} {\bibfnamefont
  {M.}~\bibnamefont {Burakowski}}, \bibinfo {author} {\bibfnamefont
  {P.}~\bibnamefont {Mrowi\'nski}}, \bibinfo {author} {\bibfnamefont
  {B.}~\bibnamefont {Krajnik}}, \bibinfo {author} {\bibfnamefont
  {M.}~\bibnamefont {Xiong}}, \bibinfo {author} {\bibfnamefont
  {K.}~\bibnamefont {Yvind}}, \bibinfo {author} {\bibfnamefont
  {N.}~\bibnamefont {Gregersen}}, \bibinfo {author} {\bibfnamefont
  {A.}~\bibnamefont {Musial}}, \bibinfo {author} {\bibfnamefont
  {A.}~\bibnamefont {Huck}}, \bibinfo {author} {\bibfnamefont {T.}~\bibnamefont
  {Heindel}}, \bibinfo {author} {\bibfnamefont {M.}~\bibnamefont {Syperek}},\
  and\ \bibinfo {author} {\bibfnamefont {E.}~\bibnamefont {Semenova}},\
  }\bibfield  {title} {\bibinfo {title} {High-throughput quantum photonic
  devices emitting indistinguishable photons in the telecom $c$-band},\
  }\href@noop {} {\bibfield  {journal} {\bibinfo  {journal} {Nature
  Communications}\ }\textbf {\bibinfo {volume} {15}},\ \bibinfo {pages} {3358}
  (\bibinfo {year} {2024})}\BibitemShut {NoStop}%
\bibitem [{\citenamefont {Galimov}\ \emph {et~al.}(2021)\citenamefont
  {Galimov}, \citenamefont {Rakhlin}, \citenamefont {Klimko}, \citenamefont
  {Zadiranov}, \citenamefont {Guseva}, \citenamefont {Troshkov},\ and\
  \citenamefont {Toropov}}]{Galimov2021}%
  \BibitemOpen
  \bibfield  {author} {\bibinfo {author} {\bibfnamefont {A.}~\bibnamefont
  {Galimov}}, \bibinfo {author} {\bibfnamefont {M.}~\bibnamefont {Rakhlin}},
  \bibinfo {author} {\bibfnamefont {G.}~\bibnamefont {Klimko}}, \bibinfo
  {author} {\bibfnamefont {Y.}~\bibnamefont {Zadiranov}}, \bibinfo {author}
  {\bibfnamefont {Y.}~\bibnamefont {Guseva}}, \bibinfo {author} {\bibfnamefont
  {T.~V.}\ \bibnamefont {Troshkov}, \bibfnamefont {S.I.~Shubina}},\ and\
  \bibinfo {author} {\bibfnamefont {A.}~\bibnamefont {Toropov}},\ }\bibfield
  {title} {\bibinfo {title} {Source of indistinguishable single photons based
  on epitaxial {I}n{A}s/{G}a{A}s quantum dots for integration in quantum
  computing schemes},\ }\href@noop {} {\bibfield  {journal} {\bibinfo
  {journal} {JETP Letters}\ }\textbf {\bibinfo {volume} {113}},\ \bibinfo
  {pages} {252–258} (\bibinfo {year} {2021})}\BibitemShut {NoStop}%
\end{thebibliography}%

\end{document}